\documentclass[graybox, secnum]{svmult}

\usepackage{mathptmx}       
\usepackage{helvet}         
\usepackage{courier}        
\usepackage{type1cm}        
\usepackage{makeidx}         
\usepackage{graphicx}        
\usepackage{multicol}        
\usepackage[bottom]{footmisc}
\usepackage{hyperref}        
\usepackage{soul}            
\hypersetup{colorlinks=true,urlcolor=blue}
\usepackage[square,numbers]{natbib}
\makeindex             

\usepackage{mathtools}
\usepackage{amsmath,amssymb,amsfonts,latexsym,stmaryrd,bbm,physics,mathrsfs}
\allowdisplaybreaks[4]

\newcommand{\be}{\begin{equation}}
\newcommand{\ee}{\end{equation}}
\newcommand{\barray}{\begin{array}}
\newcommand{\earray}{\end{array}}
\newcommand{\bea}{\begin{eqnarray}}
\newcommand{\eea}{\end{eqnarray}}
\newcommand{\bs}{\begin{subequations}}
\newcommand{\es}{\end{subequations}}

\newcommand{\bit}{\begin{itemize}}
\newcommand{\eit}{\end{itemize}}
\newcommand{\bd}{\begin{description}}
\newcommand{\ed}{\end{description}}

\def\nn{\nonumber}

\newcommand{\Id}{\mathbbm{1}}

\def\la{\langle}
\def\ra{\rangle}

\newcommand{\vet} [2] {\left ( \begin{array}{c}{#1}\\{#2} \end{array} \right ) }

\def\w{\wedge}
\newcommand{\p}{\partial}

\newcommand{\N}{\mathbb{N}}

\newcommand{\R}{\mathbb{R}}
\newcommand{\C}{\mathbb{C}}

\newcommand{\f}{\frac}
\newcommand{\tl}{\tilde}

\renewcommand{\a}{\alpha}  \newcommand{\g}{\gamma}  
\renewcommand{\d}{\delta}  \newcommand{\eps}{\epsilon} \newcommand{\z}{\zeta}
 \renewcommand{\th}{\theta}   
    \renewcommand{\l}{\lambda}
\let\m=\mu       \let\r=\rho \let\om=\omega
       
\let\G=\Gamma    \let\L=\Lambda

\newcommand{\SU}{\mathrm{SU}}
\newcommand{\SO}{\mathrm{SO}}
\newcommand{\SL}{\mathrm{SL}(2,\mathbb{C})}

\newcommand{\ut}[1]{ \underset{\widetilde{}}{#1}{} }

\newcommand{\sscr}{\scriptscriptstyle\rm}

\begin{document}

\title*{Spinfoams: Foundations${}^\star$}
\toctitle{Spinfoams: Foundations} 
\titlerunning{Spinfoams: Foundations} 
%
%

\author{Jonathan Engle${}^\dagger$ and Simone Speziale${}^\ddagger$
}
\tocauthor{Jonathan Engle and Simone Speziale} 
\authorrunning{Jonathan Engle and Simone Speziale} 
\institute{${}^\dagger$Florida Atlantic University, 777 Glades Road, Boca Raton, FL 33431, USA \\
\email{jonathan.engle@fau.edu} 
\\ 
${}^\ddagger$Aix Marseille Univ., Univ. de Toulon, CNRS, CPT, UMR 7332, 13288 Marseille, France \\ \email{simone.speziale@cpt.univ-mrs.fr}\\
\rule{0in}{0in}\\
${}^\star$Invited Chapter for the {\it Handbook of Quantum Gravity} (Eds. Bambi, Modesto and Shapiro, Springer 2023)}
\maketitle

\date{\today}
\abstract{Spinfoams provide a framework for the dynamics of loop quantum gravity that is manifestly covariant under the full four-dimensional diffeomorphism symmetry group of general relativity. In this way they complete the ideal of three-dimensional diffeomorphism covariance that consistently motivates loop quantum gravity at every step.  Specifically, spinfoam models aim to provide a projector onto, and a physical inner product on, the simultaneous kernel of all of the constraints of loop quantum gravity by means of a discretization of the gravitational path integral. In the limit of small Planck constant, they are closely related to the path integral for Regge calculus, while at the same time retaining all of the tools of a canonical quantum theory of gravity. They may also be understood as generalizations of well-understood state sum models for topological quantum field theories. In this chapter, we review all of these aspects of spinfoams, as well as review in detail the derivation of the currently most used spinfoam model, the EPRL model, calculational tools for it, and the various extensions of it in the literature. We additionally summarize some of the successes and open problems in the field. 
}

\bigskip


\tableofcontents

%
%


\newpage

\section{Introduction}

Loop quantum gravity gives a compelling description of what quantum space may look like: a collection of atoms of space, whose geometry shows all the characteristics of a quantum system, such as discrete spectra, incompatibility of classical observables, superpositions and entanglement. The crucial question for the theory is how to describe the dynamics of such quanta, in a way that is both ultraviolet complete and compatible with general relativity in the appropriate limit.
The spinfoam formalism is an approach to the dynamics based on constructing a well-defined version of the gravitational path integral with boundary states that belong to the Hilbert space of Loop Quantum Gravity. Excellent reviews and introductions to this formalism exist in the literature, see \cite{Baez:1997zt, CarloBook, PerezLR, RovelliVidotto, Bianchi:2017hjl}.
Our choice for this chapter was to present something complementary and up-to-date with respect to them.
We first go over the historical path that motivated the tools used in the formalism, starting from the idea that one should define transition amplitudes for spin networks that implement all quantum constraints in a covariant way. We then focus on a specific model, the Lorentzian EPRL model, which is both the currently most studied model, and is also rich enough to allow us to introduce all type of techniques and ideas that are also used in other models. We conclude with an overview on the current open questions and research directions.

\section{The spinfoam gist}

\subsection{The path integral as a means to implement the constraints of gravity}

Loop quantum gravity is a canonical approach to quantum gravity, with a Hilbert space of states, and operators corresponding to the basic variables of the theory. As a canonical approach, from the very start, it treats space and time very differently, a fact very much at odds with the spirit of both special and general relativity, in which space and time are part of a single unified continuum. Feynman's path integral approach opened the door to a manifestly space-time covariant formulation of quantum mechanics. It is therefore natural to ask: Is there a path integral version of loop quantum gravity, and if so what does it look like? The quest to answer this question leads to the spinfoam formalism. 

General relativity is a theory with \textit{constraints}: in addition to equations of motions which determine time evolution, there are also equations of motion that are independent of time, constraining the data at any fixed time. Such equations of motions are called constraints; when cast in the form of a set of phase space functions $C_i$ set to zero, the phase space functions $C_i$ are likewise referred to as constraints. Familiar examples of constrained systems include Maxwell theory, general relativity, and Yang-Mills theory. The theory of the quantization of constrained systems was formalized by Dirac \cite{Dirac:1958sq}. Each constraints $C_i$ whose Poisson brackets with other constraints are again constraints is called \textit{first class} and leads to a quantum equation $\hat{C}_i \psi = 0$ which supplement  Schr\"odinger's equation.  Solutions to the quantum constraints are called \textit{physical states}, and the space of all such states, equipped with an appropriate inner product, is called the \textit{physical Hilbert space}. 

The first class constraints of a theory are in one-to-one correspondence with its gauge symmetries: In order for initial data to uniquely determine a full solution to the equations of motion, it is necessary that any two phase space points related, by the flow generated by a first class constraint, be considered the same physical state. That is, the flow generated by each first class constraint must be considered a family of gauge transformations. In quantum theory, the flow generated by each first class constraint operator $\hat{C}_i$ is the family of transformations $\exp(i\lambda \hat{C}_i)$. 
When this family acts on a physical state, $\hat{C}_i\psi =0$, its action is trivial, $\exp(i\lambda \hat{C}_i)\psi = \psi$, consistent with its interpretation as a family of gauge transformations also at the quantum level. In fact, at the quantum level, the condition that $\psi$ satisfy the quantum constraint equations is equivalent to requiring that it be gauge invariant. One can distinguish between gauge transformations which preserve the polarization of the quantization from those that don't. In the first case, the transformation has a well-defined action on the argument of the wavefunction $\psi$ and acts on $\psi$ via pull-back, while in the second case the action on $\psi$ is more complicated.

The basic principle giving rise to General Relativity is the principle of general covariance, which is equivalent to covariance under the group of four dimensional diffeomorphisms, the basic symmetry group of the theory.  These both are also equivalent to Einstein's deeper principle of \textit{background independence} that there be no background space-time structure in physical law, but only dynamical space-time structures. This basic symmetry group is furthermore a gauge symmetry: Its action on initial data is exactly what is generated by the main constraints of the theory, the Hamiltonian and diffeomorphism constraints, which respectively generate time evolution with respect to an arbitrary time coordinate and spatial diffeomorphisms. 
For spatially compact regions, the Hamiltonian of the theory is furthermore a \textit{linear combination} of the constraints, so that, in quantum theory, physical states do not evolve in time --- the famous so-called `problem of time', directly related to the fact that the choice of time coordinate in general relativity is a pure gauge choice.

Just as the framework of canonical quantization needed to be generalized by Dirac, Bergmann, and others to accommodate the conceptual novelties of general relativity, so too, in order to apply Feynman's derivation of the path integral formalism to quantum gravity, it must be generalized, specifically in three different ways:

1. The path integral for an unconstrained system has the interpretation as the probability amplitude for a given initial quantum state to evolve into a final quantum state after a specified amount of time. It thereby encodes the quantum time evolution of the system --- the Schr\"odinger equation. For constrained systems, there are quantum equations of motion besides Schr\"odinger's equation, and the path integral in this case encodes them as well: The path integral between a given initial and final state provides the probability amplitude for the \textit{projection} of the initial state onto the physical Hilbert space to evolve into the final state after a given amount of time. This is true for any constrained system.
For constraints generating non-compact orbits, physical states are non-normalizable --- that is, distributional
--- and the projector onto physical states is then defined in a generalized sense.
For general relativity, because additionally quantum time evolution is trivial, no time evolution is involved in the interpretation of the the path integral, and it formally correspond to the probability amplitude for the projection of the initial state into the physical Hilbert space to be found in the final state. That is: it provides the \textit{physical inner product} between the projections of the initial and final states into the physical Hilbert space, and in this way encodes the dynamics of the theory. The new role of the gravitational path integral as projector on to physical states was noted by Hartle and Hawking \cite{Hartle:1983ai}, and later in more detail by Halliwell and Hartle \cite{Halliwell:1990qr}.

2. If one considers Feynman's derivation more carefully, one sees that, instead of integrating over classical histories, one must integrate over histories of \textit{eigenvalues} of a \textit{complete set of operators}, obtained from inserting resolutions of the identity at infinitesimal time intervals. In the case in which the spectra of the operators are continuous, this reduces to the more familiar case of an integral over classical histories. But in the case of loop quantum gravity this is not the case, and so one \textit{sums} over a certain class of \textit{discrete} space-time geometries, called spinfoams, which motivates the name for the formalism. 

3. In background-dependent field theories, path integrals provide transition amplitudes between initial and final states. This is achieved by integrating over histories in a region bounded by two Cauchy slices and assuming suitable fall-off conditions at spatial infinity. The initial and final states then belong to the Hilbert spaces associated with the initial and final slices, and if first class constraints are present, the path integral automatically projects on the physical Hilbert space, see e.g. \cite{Halliwell:1990qr,Mattei:2005cm}.
In a background-independent theory, the split into Cauchy slices has no preferred physical meaning. What one can do is to avoid it altogether, and instead apply the formalism to a \emph{general} bounded region. The result gives a quantum amplitude for the 3-geometry on this compact boundary which is formally diffeomorphism-invariant and background independent. This formulation of quantum field theory is called the \textit{general boundary} formulation. It was present in the early ideas of the Hartle-Hawking proposal and was formally developed by Oeckl \cite{Oeckl:2003vu} and applied to the LQG context starting with \cite{CarloBook,Rovelli:2005yj}, see also \cite{Speziale:2008uw} for an early review. It is central to all concrete work in spinfoams that has been done since.
The resulting quantum amplitude has a `frozen time' status, hence some interpretive work is needed to use it to describe physical processes. 
A useful procedure is to use partial observables, namely splitting the system to introduce a physical clock and re-express the amplitude 
as a function of the clock reading \cite{CarloBook}. 

\bigskip

With these generalizations, application to loop quantum gravity of Feynman's path integral leads to the spinfoam formalism. Spinfoams are not the only existing formalism for a gravitational path integral --- there are a number of precursors,  notably the Euclidean path integral formalism of Gibbons and Hawking \cite{Gibbons:1994cg}, quantum Regge calculus \cite{Rocek:1982fr}, and causal dynamical triangulations \cite{Ambjorn:2001cv, Ambjorn:2000dv, Ambjorn:1991pq}. What makes the spinfoam formalism unique is that it provides transition amplitudes (or equivalently, a projector onto physical states with physical inner product) for a canonical quantum theory of gravity with mathematically well-defined inner product and operators. In this way, the spinfoam formalism is able to overcome corresponding limitations of other approaches. At the same time, the spinfoam formalism strives to respects the basic principles of general relativity: diffeomorphism covariance or background independence, and that gravity is geometry.  

In addition, the spinfoam formalism turns out to include 
topological quantum field theories --- that is, quantum field theories with no local physical degrees of freedom, that are therefore simpler and well-understood --- that pre-date the spinfoam formalism. These early topological spinfoams are important both as toy models for understanding the four dimensional gravitational case, as well as providing building blocks for the construction of the latter, as we will see.

\subsection{Transition amplitudes for spin network states}

\label{sect:sfansatz}

Let us elaborate on the points raised above in mathematical terms. Consider first perturbative quantum gravity at the linear order, in which everything is under control, and one can see explicitly that the path integral projects on physical states. Using the flat background as reference, one can compute the path integral between an initial and a final configuration $h_{1,2}$ separated by a time interval $T$, given by
\be\label{K12}
K[h_1,h_2,T]:=\int_{h(0,\vec x)=h_1}^{h(T,\vec x)=h_2} 
{\cal D}h \, e^{i S(h)}.
\ee
The integration implements the constraints and gives a sum over energy eigenstates of the free theory \cite{Mattei:2005cm},
\be
K[h_1,h_2,T]=\sum_n e^{-i E_n T}\bar\Psi_n[h^{\sscr TT}_1]\Psi_n[h^{\sscr TT}_2],
\ee
where TT denotes the transverse-traceless parts of the spatial metric.
At the linear level all spacetime diffeomorphisms preserve the polarization, so that the meaning of gauge-invariance of the wave-functional is manifest from its argument, as discussed earlier.

Moving on to a background-independent approach, both the flat metric on the Cauchy slices, as well as the time separation $T$ on the time-like boundary, are now part of the dynamical boundary data. Let us denote by $q$ the whole classical information on the boundary. The path integral \eqref{K12} should be replaced by a formal expression like
\be
K[q]:=\int_{g|_{\p M}=q} {\cal D}g \, e^{i S(g)} .
\ee
This is the essence of the general boundary formalism: It associates a quantum amplitude to $q$ as a whole. Only after a physical clock is introduced, and an associated split of the boundary, can this be meaningfully interpreted as a transition amplitude between initial and final physical states. 
To turn this formal expression into a well-defined mathematical formula, we seek to make $K[q]$ precise such that it provides a quantum amplitude for the states of loop quantum gravity. 

Loop quantum gravity, being based on a connection formulation of general relativity, has three sets of first class constraints: The Gauss constraint, generating local gauge rotations of the $\SU(2)$ principal fiber bundle, the diffeomorphism constraint, and the Hamiltonian constraint. Because the gauge symmetries generated by the first two of these have well defined action on the connection --- the  argument of the kinematical wavefunction --- their implementation on the kinematical Hilbert space is unambiguous and straightforward. 
Recall that each normalizable state of the kinematical Hilbert space is labelled by a graph, and this corresponds to a truncation of the number of degrees of freedom. This truncation corresponds to a distributional description of the gravitational field, but can also be interpreted in terms of discrete, piecewise flat geometries, see e.g. \cite{Rovelli:2010km} and the Chapter ``Emergence of quantum Riemannian geometry'' in this Handbook \cite{Haggard:2023tnj}. Due to the truncated nature of the normalizable states, 
the gauge group generated by the quantum Gauss constraint is in a precise sense compact, so that its solutions are a subspace of the kinematical Hilbert space, with orthonormal basis given by the \textit{spin network} states $|S\rangle$.  Each of these states is labeled by an embedded graph, an assignment of an $\SU(2)$ irreducible representation to each link, and to each node an intertwiner among the representations labeling the adjacent links. 

The gauge group generated by the diffeomorphism constraint, by contrast, is non-compact, and so the solutions are distributional. A projector from spin network states to these distributional solutions, and inner product on the resulting image, is defined by averaging over the action $U(\varphi)$ of diffeomorphisms\cite{Ashtekar:1995zh}:
\begin{align}
\label{eqn:diffave}
|s(S)\rangle &:= \int {\cal D}\varphi |\varphi \cdot S\rangle, \\
\nonumber
\langle s(S'), s(S) \rangle &:= \int {\cal D}\varphi 
{\cal D}\varphi' \langle \varphi'\cdot S'| \varphi \cdot S\rangle
=  \int {\cal D}\varphi {\cal D}\varphi' \langle S'| U(\varphi'^{-1}\circ\varphi) |S\rangle \\
\label{eqn:diffprod}
&
=  \int {\cal D}\varphi \langle S'| U(\varphi) |S\rangle
= \langle S'|s(S)\rangle = \langle s(S')|S\rangle
\end{align}
where in going from the first to the second line of \eqref{eqn:diffprod}, ${\cal D}\varphi$ has been assumed to be left-invariant and have unit volume. 
The averaged spin network $s(S)$ is labeled by a spin network `up to diffeomorphism'. 
Although information about knotting of the graph survives averaging over diffeomorphisms, all known observables, as well as the dynamics considered in this chapter, are insensitive to this information, and so one can, up to some subtleties \cite{Rovelli:1995ac,Ashtekar:1995zh,CarloWinston}, think of $s(S)$ as a spin network on an abstract graph.
Note the final two expressions in \eqref{eqn:diffprod} are well-defined functions of $s(S)$ and $s(S')$, and with appropriate assumptions on the measure ${\cal D}\varphi$
\cite{Ashtekar:1995zh}, are finite. Thus, though the projection in \eqref{eqn:diffave} is not normalizable in the kinematical inner product, it \textit{is} normalizable in the inner product defined in \eqref{eqn:diffprod}.
The resulting Hilbert space is denoted $\mathcal{H}_{\text{Diff}}$.

One can then proceed to construct a projector from $\mathcal{H}_{\text{Diff}}$ to solutions of the Hamiltonian constraint again by averaging over the corresponding gauge, similar to \eqref{eqn:diffave} \cite{Reisenberger:1994aw,Rovelli:1998dx}.
The situation is however more complicated. We sketch here the procedure and its shortcomings, because it serves both as a useful review and as a motivation for spinfoams.
If $\hat{H}(x)$ denotes the Hamiltonian constraint operator, and $\hat{H}[N]$ its smearing by an arbitrary lapse $N(x)$, we have
\begin{align}
\nonumber
    &P \ket{s} := \prod_x P_x \ket{s} =  \prod_x \int_{-\infty}^{\infty} dN(x) e^{-iN(x)\hat{H}(x)} \ket{s} 
    =   \int {\cal D}N e^{-i\hat{H}[N]}\ket{s} \\
    \nonumber
    & =   \int {\cal D}\varphi \int {\cal D}N e^{-i\hat{H}[N]}\ket{\varphi \cdot S}
    = 
      \int {\cal D}\varphi \, U(\varphi)\int {\cal D}N e^{-i\hat{H}[\varphi^*N]}\ket{S} \\
    \label{hamave}
    &=   \int {\cal D}\varphi \, U(\varphi)\int {\cal D}N e^{-i\hat{H}[N]}\ket{S}
    = \sum_{m=0}^\infty \frac{(-i)^m}{m!} \int {\cal D}\varphi \, U(\varphi)\int {\cal D}N \hat{H}[N]^m\ket{S} 
\end{align} 
where $P_x$ denotes the projector onto the kernel of $\hat{H}(x)$ and $S$ is any kinematical spin network that averages to $s$. In the last step the diff-invariance of the measure ${\cal D}N$ was used. 
Note the last three expressions above are manifestly diffeomorphism invariant, in addition to formally satisfying the Hamiltonian constraint. One thus expects $P$ to project from $\mathcal{H}_{\text{Diff}}$ into solutions  of \textit{all} the constraints. 

To continue the derivation, at this point one must choose a specific proposal for the Hamiltonian constraint operator. More than one such proposal exists. In the end, as we wish to simply to motivate a general ansatz, this choice does not matter too much, and so we consider the so-called `Euclidean' term in the first one, that introduced by Thiemann in \cite{Thiemann:1996aw}.
Thiemann's smeared Hamiltonian constraint operator $\hat{H}[N]$ is defined so that it acts only at nodes of the spin network, and includes an holonomy operator that introduces a sum over spins.
The action of the Euclidean part takes the expression
\begin{align}
\label{thiemannaction}
\hat{H}[N]|S\rangle =
N(x_n)\hat{H}(x_n) \ket{S}
=N(x_n) A_n^\alpha(S) |S_n^\alpha\rangle
\end{align}
where $A_n^\a(S)$ are coefficients that can be explicitly computed, summation is understood for the repeated indices $n$ and $\alpha$, and where $n$ runs over the nodes of $S$, and, for each node $n$, $\alpha$ runs over the choice of all possible pairs of links $(\ell, \ell')$ at $n$ as well as over two irreducible representations on each of these links. $S_n^\alpha$ denotes the spin network $S$ with an extra link added connecting $\ell$ and $\ell'$ as in figure \ref{fig:hamaction}, and $x_n$ the spatial position of $n$. 
In the original prescription \cite{Thiemann:1996aw}, one lets this operator act on diffeomorphism invariant states, at which point the exact position of the added link does not matter, as long as it is chosen sufficiently close to the node (and then the limit as it approaches the node, corresponding to the removal of the regulator, becomes trivial). For the purpose of deriving the spinfoam framework, an exact position for the added link for each spin network and node therein must be chosen, and this must be done in a way that is diffeomorphism covariant, which can
always be done, e.g., by making these choices arbitrarily for one spin network in each diffeomorphism equivalence class, and then extending these choices to the rest via the action of diffeomorphisms. 

\def\freeline[#1,#2,#3,#4]{\qbezier(#1,#2)(#1,#2)(#3,#4)}
\begin{figure}[t]
\centering
\begin{minipage}[c]{1in}
\begin{center}
\setlength{\unitlength}{0.30in}
\begin{picture}(2.6,3.1)
\freeline[1,0.1,1.3,1.1]
\freeline[1.6,0.1,1.3,1.1]
\freeline[1.3,1.1, 0.1,3.1]
\freeline[1.3,1.1, 2.5, 3.1]
\put(1.215,1){$\bullet$ $n$}
\put(-0.15,2.8){$\ell$}
\put(2.55,2.8){$\ell'$}
\put(0.4,1.9){$j_\ell$}
\put(1.95,1.9){$j_{\ell'}$}
\end{picture}
\end{center}
\end{minipage}
\begin{minipage}[c]{0.6in}
\begin{center}
\setlength{\unitlength}{1.9in}
\begin{picture}(0.25,0.05)
\put(0,0){\vector(1,0){0.25}}
\end{picture}
\end{center}
\end{minipage}
\begin{minipage}[c]{1in}
\begin{center}
\setlength{\unitlength}{0.30in}
\begin{picture}(2.6,3.1)
\freeline[1,0.1,1.3,1.1]
\freeline[1.6,0.1,1.3,1.1]
\freeline[1.3,1.1,0.1,3.1]
\freeline[1.3,1.1,2.5,3.1]
\freeline[0.7,2.1,1.9, 2.1]
\put(1.215,1){$\bullet$ $n$}
\put(-0.15,2.8){$\ell$}
\put(2.55,2.8){$\ell'$}
\put(0.125,2.35){$j_\ell$}
\put(2.25,2.35){$j_{\ell'}$}
\put(1.1875,2.35){$\frac{1}{2}$}
\put(-0.1,1.45){$j_\ell \pm \frac{1}{2}$}
\put(1.7,1.45){$j_{\ell'} \pm \frac{1}{2}$}
\end{picture}
\end{center}
\end{minipage}
\caption{\label{fig:hamaction} 
Action of Thiemann's Hamiltonian constraint operator on a node $n$ of a spin network.}
\end{figure}
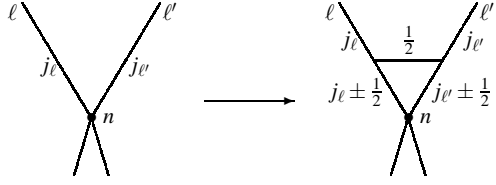
For repeated actions of $\eqref{thiemannaction}$
on a spin network, it is convenient to introduce the notation 
$S_{n_1 \cdots n_m}^{\alpha_1\cdots \alpha_m}:= (\cdots(S_{n_1}^{\alpha_1}) \cdots)_{n_m}^{\alpha_m}$,
so that, for example,
\begin{align}
\hat H[N]^2 \ket{S}
= \hat H[H] 
\left(N(x_{n_1}) A_{n_1}^{\alpha_1}(S) |S_{n_1}^{\alpha_1}\rangle\right)
= N(x_{n_2}) N(x_{n_1}) 
A_{n_2}^{\alpha_2}(S) A_{n_1}^{\alpha_1}(S) |S_{n_1 n_2}^{\alpha_1 \alpha_2}\rangle
\end{align}
With this notation, continuing the derivation \eqref{hamave},
\newcommand{\tightcdots}{\!\cdot\!\!\cdot\!\!\cdot\!}
\begin{align}
\nonumber
 &P \ket{s} \! =  
\!\!
\sum_{m=0}^\infty \! \frac{(-i)^m}{m!} \int\!\! {\cal D}\varphi \, U(\varphi) \! \int \!\! {\cal D}N \, 
 N(x_{n_m}) \tightcdots N(x_{n_1}) A_{n_m}^{\alpha_m}(S_{n_1 \cdots n_{m-1}}^{\alpha_1\cdots \alpha_{m-1}}) \tightcdots A_{n_1}^{\alpha_1}(S) 
 \ket{S_{n_1 \cdots n_m}^{\alpha_1\cdots \alpha_m}} \\
 \label{hamavefinal}
&
=  \sum_{m=0}^\infty \frac{(-i)^m}{m!} \left(\int {\cal D}N 
\, N(x_{n_m}) \tightcdots N(x_{n_1})\right) A_{n_m}^{\alpha_m}(S_{n_1 \cdots n_{m-1}}^{\alpha_1\cdots \alpha_{m-1}}) \tightcdots A_{n_1}^{\alpha_1}(S) \ket{s_{n_1 \cdots n_m}^{\alpha_1\cdots \alpha_m}}
\end{align}
where $s_{n_1 \cdots n_m}^{\alpha_1\cdots \alpha_m}:= s(S_{n_1 \cdots n_m}^{\alpha_1\cdots \alpha_m}:= (\cdots(S_{n_1}^{\alpha_1}) \cdots)_{n_m}^{\alpha_m})$.
The diff-invariance of ${\cal D}N$ further ensures that the remaining (formal) integral over the lapse,
\begin{align} 
I_{\{m_k\}}:=\int {\cal D}N \, N(x_{n_m}) \cdots N(x_{n_1}),
\end{align}
depends only on the number $m_k$ of points that appear $k$ times for $k=1,2,3,\dots$.
Using the projection \eqref{hamavefinal} to define the physical inner product on 
solutions to the Hamiltonian constraint,
in a manner similar to \eqref{eqn:diffprod}, we have
\begin{align}
\label{eq:amp2}
 \bra{s'} P \ket{s}
=  \sum_{m=0}^\infty \frac{(-i)^m}{m!} I_{\{m_k\}} 
A_{n_m}^{\alpha_m}(S_{n_1 \cdots n_{m-1}}^{\alpha_1\cdots \alpha_{m-1}}) \cdots A_{n_1}^{\alpha_1}(S)
\langle s'\ket{s_{n_1 \cdots n_m}^{\alpha_1\cdots \alpha_m}} .
\end{align}
This expression can be interpreted as a sum over sequences of diffeomorphism-invariant spin networks $(s, s_{n_1}^{\alpha_1}, \dots, s_{n_1 \cdots n_m}^{\alpha_1\cdots \alpha_m})$ related by consecutive actions of the Hamiltonian constraint, acting as in figure \ref{fig:hamaction}, and ending in the final spin network $s'$. Such a sequence of spin networks can be interpreted as a \textit{spinfoam}, a quantum space-time, matching to the initial and final spin networks $s$ and $s'$, which have the interpretation of states of quantum ``space''. Each such spinfoam consists in a 2-complex, with each 2-cell, or face $f$, labelled by a quantum number of area, $j_f$, and each 1-cell, or edge $e$, labelled by an intertwiner $i_e$. An edge branches into multiple edges at a \textit{vertex} $v$, corresponding to one of the actions of the Hamiltonian constraint at a single node. These vertices $v$ are thus in one to one correspondence with the factors $A_{i_m}^{\alpha_m}(S_{n_1 \cdots n_{m-1}}^{\alpha_1\cdots \alpha_{m-1}})$ appearing in the above expression. Each such factor depends only on the spins and intertwiners at the node acted upon and the nodes created, and, due to the diffeomorphism covariance of $\hat{H}[N]$, is independent of the choice of representative $S$ averaging to $s$. We therefore denote each such factor as $A_v$, the vertex amplitude associated to $v$. The amplitude \eqref{eq:amp2} can then be written
\begin{align}
\langle s' \mid P \mid s\rangle =  \sum_{\mathcal{F}|\partial \mathcal{F} = s\cup s'} \frac{(-i)^m}{m!}I_{\{m_k\}} \prod_{v\in \mathcal{F}} A_v(\mathcal{F}),
\end{align}
where the sum over $\mathcal{F}$ is a sum over all spinfoams with boundary equalling the union of the initial and final spin networks $s,s'$, $m$ is the number of vertices in $\mathcal{F}$, and $m_k$ the number of nodes in $s'$ 
with $k$ vertices in $\mathcal{F}$ resulting from the action of $\hat{H}[N]$ on it. 

The resulting expression is however still unsatisfactory. 
The vertex amplitude $A_v(\mathcal{F})$ depends on how the spinfoam $\mathcal{F}$ is interpreted as a history of spin networks, which can be related to a choice of foliation of spacetime. Furthermore, the summand, due to the factor $I_{\{m_k\}}$, is not local in time.
In addition, it is not yet clear how to make sense of the factors $I_{\{m_k\}}$ resulting from a functional integral over lapses \cite{Rovelli:1998dx}. Therefore, even though the construction of a well-defined Hamiltonian operator was a key feature to support the viability of the program, there are unfortunately ambiguities in its definition, and it is not even clear that it satisfies all necessary physical requirements.
The idea to overcome these limitations is to take the above expression as a suggestion for a more general ansatz, which keeps the key state-sum structure, and makes foliation-independence manifest:
\begin{align}
\label{sfansatz}
K[s' \cup s] :=
\langle s' \mid P \mid s\rangle = \sum_{\mathcal{F}|\p\mathcal{F}=s\cup s'} A_\mathcal{F}, \qquad 
A_\mathcal{F}=\prod_f A_f \prod_e A_e\prod_v A_v,
\end{align}
where each $A_f$ depends only on the spin label $j_f$ on $f$, $A_e$ depends only on the intertwiner $i_e$ on $e$, and $A_v$ depends only on the spins and intertwiners on the faces and edges incident at the vertex $v$. The goal is to find expressions for the weights such that the quantum constraints are correctly implemented, and general relativity is recovered in the appropriate limit.

Further support for a formula like \eqref{sfansatz} comes from the fact that different and independent approaches have converged to it, suggesting that an expression of this type may be seen as a definition of a generally covariant QFT \cite{CarloBook}.
 State sum models for topological quantum field theories, such as the Ooguri and Crane-Yetter models for BF theory \cite{Ooguri:1992eb,Crane:1993if}, and the Ponzano-Regge and Turaev-Viro models for 3D gravity \cite{Ponzano:1968,Turaev:1992hq}, take this form, as do group field theories, appropriately cast \cite{Oriti:2017ave}.
 It is referred to as the \textit{spinfoam ansatz}, 
 and it forms the basis of the spinfoam
framework for defining dynamics for loop quantum gravity.
Requiring the composition property
\begin{align}
\sum_{s''} K[s' \cup s''] K[s''\cup s]
= \sum_{s''}\bra{s'} P \ket{s''} \bra{s''} P \ket{s}
= \bra{s'} P \ket{s} = K[s'\cup s]
\end{align}
fixes the face amplitudes to be $A_f = 2j_f+1$ \cite{Carlodimj}. Furthermore, the edge amplitude can always be absorbed into the vertex amplitude, so that the latter carries all the dynamical information of the model. 

The formula \eqref{sfansatz} has the same structure as a partition function for lattice gauge theory (LGT), and some techniques for its study are indeed common to the two fields. However, there are also crucial differences stemming from the differences between quantizing a gauge theory on Minkowski spacetime versus quantizing general relativity with no background structures. 
In particular, LGT can use a regular lattice adapted to flat spacetime, and this is pivotal in order to derive the right universal class of actions, define observables and length scales, compute the continuum limit and verify the existence of a phase transition. 
Conversely, background independence means that one can only use abstract graphs, whose regularity and locality are a priori distinct from that of any emerging semiclassical spacetime, so that verifying that the action is related to gravitational dynamics becomes subtle and non-trivial, and there is as yet no consensus on a procedure to define the relevant scales and corresponding choice of boundary and bulk structure to compute specific physical processes. Similarly, the refinement limit and how to sum over refined contributions are open questions, whose study is the object of ongoing research.
See in particular the Chapter ``Spin foams, Refinement limit and Renormalization'' in this Handbook dedicated to these questions \cite{Asante:2022dnj}. The recasting of spin foams as group field theories also provides an avenue for answering these questions \cite{Oriti:2014yla}.  As we will comment on in the conclusions, more explicit results are needed before one can discern between different procedures put forward in the literature.
On a more technical level, the LGT dynamics is encoded in the face amplitude via the plaquette action and the vertex amplitude carries only information about gauge-invariance, whereas in most spinfoam models it is the reverse: 
It is the vertex amplitude that encodes dynamics, and the face amplitude is determined by gauge invariance with no dynamical information.
The similarities and differences between LGT and spinfoam quantum gravity are a rich topic for discussions, see e.g. \cite{Rovelli:2011fk}.

\subsection{Spinfoam amplitudes from the Plebanski action}

As noted, an expression like \eqref{sfansatz} is known to arise in quantum BF theories; furthermore, in 4d, the action for general relativity can be written as a constrained BF action. This suggests the following approach: first quantize BF theory, thus obtaining an amplitude already in the form  \eqref{sfansatz}, then impose constraints reducing BF to gravity directly at the quantum level, as a restriction on the quantum states allowed. There is no guarantee that this procedure gives the same result as quantizing the constrained theory, if one knew how to do that. The logic used is that one will have to verify a posteriori whether one obtains in this way a theory with the right semiclassical dynamics, and capable of making new predictions. The value of this approach is that one obtains in a natural way both a background independent quantum path integral, and boundary states given by $\SU(2)$ spin networks. There exists excellent overviews of this approach, e.g. \cite{PerezLR}. We refer the reader there for details, and only sketch here the steps necessary to follow the rest of the chapter.

The canonical structure of LQG can be derived from the tetrad action for general relativity. An equivalent way of studying the same theory is to use a version of the Plebanski action \cite{Plebanski,Reisenberger:1998fk,DePietri:1998hnx}, which is given by\footnote{The version used is often called covariant or non-chiral, to distinguish it from the original Plebanski action, which is identical but uses self-dual variables only, and has to be supplemented with reality conditions. 
}
\be
\label{S}
S(B^{IJ},\om^{IJ},\phi_{IJKL}) = \f1{16\pi G}\int \tr \left( P_\g \, B \w F(\om) - \f12 \left(\phi +\f\L3  P_\g \right) B \w B\right),
\ee
where $\om^{IJ}$ is an $\SO(3,1)$ connection with curvature $F$, $B^{IJ}$ an auxiliary 2-form valued in the adjoint, and
\be
P_\g=\f{\Id}{\g}+\star, \qquad P^{-1}_\g = \f{\g}{1+\g^2}(\Id -\g\star).
\ee
The trace, $\tr$, here means contracting all indices with the identity $\Id_{IJKL}:=\eta_{I[K}\eta_{L]J}$, and $\star:=\tfrac12\eps_{IJKL}$.
The field $\phi^{IJKL}$ is a Lagrange multiplier with the same symmetries as the Riemann tensor, and hence has 20 independent components. 
Its variation gives the following equations,
\be\label{simpl}
B^{IJ}\w B^{KL} = \f1{12} \eps^{IJKL} \,  \tr(B\w\star B).
\ee
Solutions satisfying the non-degeneracy condition $ \tr(B\w\star B) \neq 0$ are parametrized by the existence of a non-degenerate tetrad $e_\m^I$ such that one of the following four forms hold \cite{Reisenberger:1998fk,DePietri:1998hnx,Buffenoir:2004vx}
\begin{align}\label{sectors}
(I\pm) \quad 
B^{IJ}=\pm e^I\w e^J, \qquad
(II\pm) \quad 
B^{IJ}=\pm \star e^I\w e^J.
\end{align}A 2-form written as the wedge product of two 1-forms is called simple, and \eqref{simpl} are called (covariant) simplicity constraints.
Plugging in the form $(I\pm)$ recovers the original tetrad formulation with Newton constant $\pm G$ and Barbero-Immirzi parameter $\g$.
Plugging in the form $(II\pm)$ recovers the formulation with Newton constant $\pm \gamma G$ and Barbero-Immirzi parameter 
$-1/\g$. At the classical level, there is no problem in choosing either sector for any finite value of $\g$, provided the right Newton constant is identified. At the quantum level however, these two sector could lead to non-trivial interference. The linear reformulation of the simplicity constraints used in the EPRL model, which we introduce below \eqref{linsimp}, fortunately eliminates the sectors $(II\pm)$.

The covariant Plebanski action can also be seen as a theory for the two Urbantke metrics corresponding to self-dual and antiself-dual sectors, and the simplicity constraints as imposing the matching of the two metrics \cite{Speziale:2010cf, Beke:2011mu}. This approach is relevant for recent proposals to construct effective spinfoam theories \cite{Borissova:2022clg}.

The canonical structure of the theory is the following. After 3+1 splitting, the phase space is described by the conjugated pair $(\om_a^{IJ},\tl P^a_{IJ})$ and has 36 dimensions per space point. Here $\tl P^{aIJ}:= \tfrac12\tl\eps^{abc}J^{IJ}_{bc}$, and $J:=P_\g\, B/16\pi G$, and the index $a=1,2,3$ is over each leaf of the foliation.
The momenta are not all independent, and have to satisfy six primary simplicity constraints which are the canonical version of the quadratic covariant constraints \eqref{simpl}. These are not stable under evolution, and give rise to six secondary simplicity constraints, which coincide with the $D_{[a} e^0{}_{b]}=0$ plus three of the $D_{[a} e^i{}_{b]}=0$ torsionless equations (the remaining six give the Gauss constraints). Together, the primary and secondary simplicity constraints form a second class system. This together with the six Gauss constraints and the four diffeomorphism constraints reduce the phase space to 4 dimensions per space point, namely the theory has two degrees of freedom per point.

It is often convenient to fix one of the tetrad forms, say $e^0$, in terms of the hypersurface normal. This breaks part of the internal gauge symmetry, and introduces three additional constraints which form a second class pair with the generators of the broken internal symmetry.\footnote{Namely the boosts in the standard case of a space-like 3+1 splitting. Working with this partial gauge-fixing is also convenient if one wants to describe time-like or null foliations \cite{Alexandrov:2014rta}.} The three new constraints are linear \cite{Engle:2007uq,Engle:2007wy,Alexandrov:2007pq,Wieland:2010ec},
\be
\label{linsimp}
N_I (\Id-\g\star)J^{IJ}_{ab} = 0,
\ee
and imply, and so can replace, the primary simplicity constraints \eqref{simpl}. 
These are 9 independent constraints per point. The counting now has $9+9$ second class and $3+4$ first class constraints, giving 4 dimensions per space point as before.

Let us neglect for a moment the spatial indices $ab$ (imagine for instance integrating the 2-form on a given surface). If we choose the canonical time direction $N^I=t^I:=(1,0,0,0)$, this equation is equivalent to 
\be\label{KgL}
\vec K=\g \vec L,
\ee
where $K^{i}:=J^{0i}$, $L_i:=-\tfrac12\eps_{ijk}J^{jk}$. For an arbitrary direction $N^I:=\L^I{}_J t^J$, we have instead \cite{Dona:2020xzv}
\begin{align}\label{gsboosted}
\cosh \eta ( \vec K-\g  \vec L) = \sinh \eta \, (R^{-1} \vec u) \times ( \vec L+\g \vec K),
\end{align}
where the Lorentz transformation $\L$ has been decomposed as a product of a rotation $R$ times a boost with rapidity $\eta$ and direction $u$. While $B$ is simple, we refer to $J$ satisfying \eqref{KgL} as $\g$-simple, and \eqref{gsboosted} as boosted $\g$-simple.

The relevance of this action for us is that it formulates general relativity as a topological field theory plus constraints. A topological field theory can be quantized without any reference to a background metric structure, in two steps. First, invoking topological invariance to restrict the partition function/generating functional to a cellular decomposition. Second,  choosing to represent it as a product of delta functions imposing flatness of the holonomies on the faces of this decomposition. Gauge divergences are treated by removing redundant delta functions. The result is finite and can be naturally recast as a state sum model using the character expansion of the deltas. Integrating out gives precisely an expression like \eqref{sfansatz}, with the amplitudes obtained from the recoupling theory of the chosen gauge group.
This quantization of BF theory will be reviewed in details in the next Section.

It then leads to the following strategy to construct spinfoam models: start from the quantum BF theory and impose the primary simplicity constraints, which have a simple action, and can be done with a restriction on the quantum labels of the state sum model. This restriction destroys the proof of topological invariance, therefore the resulting model could carry local degrees of freedom in the continuum limit. Furthermore, the limit of large label is related to the Regge action, which offer a possibility towards proving that the continuum limit is related to GR.

Different ways of implementing this strategy have been considered in the literature. One has to first choose a discretization of the classical constraints adapted to the cellular decomposition used, and then choose a quantization map. The simplest choice of cellular decomposition is a simplicial one: In this case a piece-wise flat geometry can be unambiguously described by the edge lengths, and this description forms the basis of Regge calculus. The more adaptable choice is to work with all tetrahedra being space-like. This is the case most used in the literature. We will comment in Section~\ref{othersig} below on other choices. On each 4-simplex, the quadratic covariant constraints \eqref{simpl} can be discretized integrating the $B$ field on different triangles of the 4-simplex. The resulting constraints 
 can be split in three groups \cite{Baez:1999tk}, diagonal, cross-diagonal, and opposite-faces. Gauge invariance can be implemented in the form of closure constraints.
Flatness of the 4-simplex and closures then imply that the opposite-faces constraints are redundant and can be discarded. Indeed, the diagonal and cross-diagonal together with the closure constraints are enough to guarantee that the bivectors discretizing the $B$ field are in 1-to-1 correspondence with an oriented Regge 4-simplex. Since the opposite-faces constraints are the only ones to involve the connection, these two results could be taken as signal that imposing the primary constraints at all times, one is effectively inducing the secondary ones as well. 
 This property, however, is lost if one uses a more general cellular decomposition, because in this case the opposite-face constraints become non-trivial, and furthermore there is no unique map between edge lengths and piece-wise flat metrics. 
Finally, the closure constraints also guarantee that the diagonal and cross-diagonal are solved when linear simplicity is solved \cite{Engle:2007qf}. Hence one can trade those quadratic constraints for the linear ones, with the advantage of removing already at the classical level the presence of the  sectors $(II\pm)$ \eqref{sectors}.

Within this classical set-up, two different quantization maps have been studied.
The first one was the Barrett-Crane (BC) model \cite{Barrett:1997gw,Reisenberger:1998bn,Baez:1999tk,Barrett:1999qw}, 
later followed by the Engle-Pereira-Rovelli-Livine (EPRL) model \cite{Engle:2007uq,Engle:2007qf,Engle:2007wy,Freidel:2007py,Livine:2007ya}.
We choose to focus here on the EPRL model for two reasons. First, it is currently favoured because it overcame difficulties with correlators \cite{Alesci:2007tx}, because it implements weakly the non-commuting constraints \cite{Engle:2007uq,Livine:2007ya}, can accomodate an arbitrary Immirzi parameter \cite{Engle:2007wy,Freidel:2007py}, and can be easily extended to arbitrary graphs \cite{Kaminski:2009fm,Ding:2010fw}. Second, it is sufficiently general and versatile that it can be used as a starting point to then learn the simpler BC model, or the more complicated models with building blocks of different spacetime signatures.
In the next Section, we present in some details the EPRL model. For other work on the BC model, see e.g.\cite{Christensen:2009bi,Baratin:2011tx,Kaminski:2013yca,Jercher:2021bie,Jercher:2022mky}.

\section{EPRL model}

The EPRL model is based on the Lie group $\SL$, which describes the local Lorentz symmetry of general relativity.
In this Section we briefly review relevant aspects of this group, and explain how they are used to define the EPRL model and its relation to topological BF theory. We then review the key properties of the model, starting from its large spin limit, as well as various extensions thereof that have been studied.

\subsection{Unitary irreps of $\SL$}

The EPRL model is based on the principal series of the unitary irreducible representations (irreps) of $\SL$, which are infinite dimensional since the group is non-compact.
These can be labelled by a pair $(\r,k)\in(\R,\N/2)$, with Casimirs
\be
\hat{C}_1:={\hat {\vec L}}{}^2-{\hat {\vec K}}{}^2 =( k^2 - \rho^2 -1 )\Id, \qquad  \hat{C}_2:={\hat {\vec K}}\cdot  {\hat {\vec L}}= \rho k\Id. 
\ee
Here $(\hat {\vec L},\hat {\vec K})$ are the rotation and boost generators, as defined by the canonical time direction $t^I:=(1,0,0,0)$.
A basis for these irreps can be obtained looking at representations of any little group. For the EPRL model it is convenient to work with Naimark's canonical orthonormal basis, labelled by the $\SU(2)$ little group with $\hat {\vec L}{}^2$ and $\hat L_z$: $\ket{\r,k;j,m}$, where $j\geq k$, and $m\in[-j,j]$ is the magnetic number. The matrix elements in this basis are
$D^{(\r,k)}_{jmln}(h)$, where $jm$ and $ln$ refer respectively to the right-invariant and left-invariant realizations of the little group. If $h=g\in\SU(2)$, then 
$D^{(\r,k)}_{jmln}(g)=\d^{jl} D^{(j)}_{mn}(g)$, for all $\r$ and $k\leq j$, where $D^{(j)}_{mn}(g)$ are the usual Wigner D-matrices for $\SU(2)$.

The label $m$ of the orthonormal basis can be given a geometric interpretation as projection along the $\hat z$ axis of the rotation generator $\hat L^i$. As such, each state has minimal information about the direction of the rotation generator. A sharper characterization can be obtained working with $\SU(2)$ coherent states, which can be embedded in the unitary irreps of $\SL$ as follows,
 \be\label{SL2Ccs}
\ket{\r,k;j,\z}:=D^{(j)}\big(\z\big)\ket{\r,k;j,-j}.
\ee
Here $D(\z)$ is the Hopf section, and the magnetic index is replaced by a complex number $\z$ which represents a point on the sphere via stereographic projection. We will denote $n(\z)$ the corresponding unit vector in $\R^3$. 
The expectation values of rotations and boosts point in the direction identified by the state, 
\begin{subequations}
\label{LKev}
\begin{align}
&\vec L:= \bra{\r,k;j,\z} \hat {\vec L} \ket{ \r,k;j,\z} = -jn(\z), \\  &\vec K:= \bra{\r,k;j,\z}  {\hat{\vec K}} \ket{ \r,k;j,\z} = -\f{\r k}{j+1} n(\z).
\end{align}
\end{subequations}
The uncertainty is minimized in the direction of the rotation generators -- but not of the boost generators.\footnote{This $\SL$ embedding of $\SU(2)$ coherent states should not be confused with the $\SL$ coherent states defined in \cite{Perelomov}, which exist only for $k=0$.} 
From this relation it follows that
\be\label{KgenL}
\vec K- \f{\r k}{j(j+1)} \vec L=0,
\ee

\subsection{Classical and Quantum BF theory in four dimensions}
\label{sect:bf}

\makeatletter
\def\smallunderarrow@#1#2#3#4{%
	\vtop{\ialign{##\crcr
		\setbox0=\hbox{$\scriptscriptstyle #1$}%
		\hskip0.5\wd0$\m@th\hfil#3#4$\crcr
		\noalign{\nointerlineskip\kern1.3\ex@}%
		$\scriptscriptstyle #1$\ifx#3\displaystyle\scriptsize\else\tiny\fi#2#3\crcr}}}
\newcommand\pback[1][\relax]{\mathpalette{\smallunderarrow@{#1}\leftarrowfill@}}
\newcommand\ppback[1][\relax]{\mathpalette{\smallunderarrow@{#1}\Leftarrowfill@}}
\makeatother

The basic variables of $\SL$ BF theory on a four dimensional manifold $M$ are an $\SL$-connection $\omega^{IJ}$ and two form $J^{IJ}$ taking values in the Lie algebra 
$\mathfrak{so}(3,1) \cong \mathfrak{sl}(2,\mathbb{C})$, 
and the action is given by
\begin{align}
S_{BF}(\omega,J) = \int_M \tr (J \wedge F)
\end{align}
where $F^{IJ}$ is the curvature of $\omega^{IJ}$. Variation gives
\begin{align}
\delta S_{BF}(\omega,J) = \int_M \tr (\delta J \wedge F - (d_\omega J)\wedge \delta \omega) + \int_{\partial M} \tr (J\wedge \omega)
\end{align}
from which one can read off the equations of motion
\begin{align}
\label{bfflat}
F &\approx 0 \quad \text{(flatness)}, \\
\label{bfconst}
d_\omega J &\approx 0 \quad \text{(covariantly constant $J$)},
\end{align}
and symplectic potential $\Theta(\delta) = \int_{\partial M} J\wedge \delta \omega$, yielding a boundary phase space parameterized by 
the pull-backs $(\pback{J}, \pback{\omega})$ of the basic variables to the boundary, with Poisson brackets
\begin{align}
\label{eq:contpb}
\{J_{ab}^{IJ}(x), \omega_{c}^{KL}(y)\} = \ut\epsilon_{abc} \Id^{IJKL} \delta^{(3)}(x,y) 
\end{align}
with $\ut\epsilon_{abc}$ the Levi-Civita symbol of density weight $-1$ on $\partial M$.
Holding $\pback{\omega}$ constant on the boundary, the (formal) continuum path integral is then
\begin{align}
\label{eq:contpi}
W(\pback{\omega}) &= \int {\cal D}\omega {\cal D}J e^{i \int_M \tr(J \wedge F)} = \int {\cal D}\omega \prod_{x\in M} \delta(F)
\end{align}
which manifestly implements the flatness constraint \eqref{bfflat}.
Introduce an oriented cell complex $\mathcal{K}$ covering $M$. $\omega$ is naturally discretized by the parallel transport 
$h_e \in \SL$ it defines along each interior 1-cell, that is \textit{edge}, $e$ in $\mathcal{K}$.
The edges in the boundary $\partial \mathcal{K}$ we call \textit{links} $\ell$, which together form the \textit{boundary graph} $\gamma$. The corresponding parallel transports $h_\ell$
provide the natural discretization of $\pback{\omega}$.
The (formal) continuum path integral \eqref{eq:contpi} then has the obvious discretization
\begin{align}
\label{discpi}
W(\{h_\ell\}) & = \int \left(\prod_{e \in \mathcal{K}} dh_e\right) \prod_{f \in \mathcal{K}} 
\delta\left(\overrightarrow{\prod}_{e \in f}h_e\right).
\end{align}
where the product over $f$ is the the product over all interior 2-cells, or \textit{faces}, in $\mathcal{K}$, and the product inside of the delta function is over the 1-cells in the boundary of $f$, whether interior or boundary (link). $dh_e$ denotes the Haar measure, and $\delta(\cdot)$ denotes the Dirac delta with respect to this measure, peaked at the identity.

The group elements $h_\ell$ associated to the boundary graph $\gamma$ parameterize the discrete boundary configuration space $\mathcal{C}$, so that $\mathcal{C} \cong \SL^{L}$, where $L$ is the number of links. To discretize the conjugate variable $\pback{J}$, for each link $\ell$ of $\gamma$, introduce an oriented 2-surface $S_{\ell}$ containing the `source' node $\ell_-$ of $\ell$, with $\ell$ `above' $S_\ell$ and all other links at $\ell_-$ `below' $S_\ell$, and define the \textit{flux} $J_\ell^{IJ}:= \int_{S_\ell} J^{IJ} \in \mathfrak{so}(3,1) \cong \mathfrak{sl}(2,\mathbb{C})$. Then \eqref{eq:contpb}
yields the Poisson brackets \cite{Engle:2007mu}
\begin{align}
\label{eq:discpb}
\{J_\ell^{IJ}, h_{\ell'}  \} = \d_{\ell\ell'}h_{\ell} \tau^{IJ}
\end{align}
where $\tau^{IJ} = -\tau^{JI} \in \mathfrak{sl}(2,\mathbb{C})$ is given by $\tau^{0i}:=\frac{1}{2}\sigma_i$ and 
$\tau^{ij}:= \frac{i}{2} \epsilon^{ikj} \sigma_k$ with $\sigma_i$ the usual Pauli matrices. 
All components of each $h_\ell$ trivially Poisson commute with each other, which, together with \eqref{eq:discpb} and the Jacobi identity, yields the further non-trivial Poisson bracket \cite{Ashtekar:1998ak}
\begin{align}
\label{Jpb}
\{J_\ell^{IJ}, J_{\ell'}^{KL}\} = \d_{\ell\ell'}C^{IJKL}{}_{MN} J_{\ell}^{MN},
\end{align}
where $C^{IJKL}{}_{MN} $ are the structure constants of $\mathfrak{so(3,1)}$. All other Poisson brackets are zero.
These Poisson brackets can also be derived from a discrete action \cite{EPRlong,Dittrich:2008ar}.
They give the discrete boundary phase space $\Gamma := \{(h_\ell, J_\ell)_{\ell \in \gamma}\}$ the structure of the cotangent bundle over $\mathcal{C} \cong \SL^L$, 
\begin{align} 
\label{cotbundle}
T^* \SL^L := \{(h_\ell, \mu_\ell)_{\ell \in \gamma} \,\, | \,\, \mu_\ell \in T^*_{h_\ell} \SL \} 
\end{align}
in which $J_\ell^{IJ}$ is identified with $\langle \mu_\ell, L_{\tau^{IJ}} \rangle$, where $L_x$ denotes the left invariant vector field associated to the algebra element $x$. 
Quantization yields the boundary Hilbert space
\begin{align}
\mathcal{H}_{\partial F} = L_2(\SL^L)
\end{align}
with basic operators $\left(\hat{h}_\ell\right)^A{}_B$ acting by multiplication, and $\hat{J}_\ell^{IJ}$ 
as the derivative operator corresponding to $L_{\tau^{IJ}}$ on the copy of the group corresponding to $\ell$. The commutators of the basic operators mimic exactly the Poisson brackets of their classical counter parts. In particular, this means that, as in loop quantum gravity, the fluxes do not commute, but rather satisfy the same algebra as the gauge group, $\mathfrak{sl}(2,\mathbb{C})$. 

If we interpret the equation of motion \eqref{bfconst} to be a priori satisfied along the interior of each link $\ell$ (but not at its end points), then, in addition to the flux $J_\ell$ at the source node of $\ell$, we can define the flux at the target node of $\ell$ as the parallel transport of $J_\ell$ via $h_\ell$. That is, for $n$ the source of $\ell$ and $n'$ the target of $\ell$, define
\begin{align}
\label{secondflux}
J_\ell(n) := J_\ell, \qquad J_\ell(n'):= h_\ell \triangleright J_\ell
\end{align}
where $\triangleright$ denotes the adjoint action. In terms of the cotangent bundle structure of the phase space  \eqref{cotbundle},  $J_\ell(n')$ then takes the form $\langle \mu_\ell, R_{\tau^{IJ}} \rangle$ with $R_x$ the right invariant vector field corresponding to $x$. 
Quantization of the expression for $J_\ell(n')$ in \eqref{secondflux} then leads to $\hat{J}_\ell^{IJ}(n')$ acting as the derivative operator corresponding to $R_{\tau^{IJ}}$ on the copy of $\SL$ corresponding to $h_\ell$. 

For each link $\ell$ and node $n$ adjacent, the operators $\hat{J}_\ell^{IJ}(n)$ satisfy the Lorentz algebra, with rotation and boost generators given by 
\begin{align}
\hat{L}_\ell^i(n) := -\frac{1}{2}\epsilon^i{}_{jk}\hat{J}_\ell^{jk}(n), \qquad \hat{K}_\ell^i(n) := \hat{J}_\ell^{0i}(n).
\end{align}
For a given link $\ell$, there are two sets of the above generators, one for each node $n,n'$, acting as the left and right invariant vector fields on the corresponding group element $h_\ell$. The $jm$ and $ln$ subscripts of the $\SL$ D-matrices of the last subsection, evaluated for $h=h_\ell$, correspond to these two sets of generators.
From the previous section, the corresponding Casimir operators are 
\begin{align}
\hat{C}_{1, \ell}:= \hat{L}_\ell(n)^2 -  \hat{K}_\ell(n)^2, \qquad \hat{C}_{2,\ell}:= \hat{L}_\ell(n)\cdot  \hat{K}_\ell(n).
\end{align}
Since the fluxes on either side of a given link $\ell$ are related by parallel transport, they yield the same Casimirs, so that the above operators are independent of $n$.  The simultaneous eigenvalues of the above operators, as well as $\hat{L}_\ell(n)^2$ and $\hat{L}_\ell(n)_z$, are then parameterized by the set of quantum numbers $(\rho_\ell, k_\ell, j_{\ell n}, m_{\ell n})$.  For each such set of quantum numbers for each $\ell$ and $n\in \ell$, there is a state $\psi_{\{\rho_\ell, k_\ell, j_{\ell n}, m_{\ell n}\}}$ in the boundary Hilbert space, unique up to phase, satisfying the equations
\begin{align}
\nonumber
\hat{C}_{1, \ell} \psi_{\{\rho_\ell, k_\ell, j_{\ell n}, m_{\ell n}\}} &= (k_\ell^2-\rho_\ell^2-1) \psi_{\{\rho_\ell, k_\ell, j_{\ell n}, m_{\ell n}\}} \\
\nonumber
\hat{C}_{2,\ell} \psi_{\{\rho_\ell, k_\ell, j_{\ell n}, m_{\ell n}\}} &= \rho_\ell k_\ell \psi_{\{\rho_\ell, k_\ell, j_{\ell n}, m_{\ell n}\}} \\
\nonumber
\hat{L}_\ell(n)^2 \psi_{\{\rho_\ell, k_\ell, j_{\ell n}, m_{\ell n}\}} &=  j_{\ell n}( j_{\ell n}+1) \psi_{\{\rho_\ell, k_\ell, j_{\ell n}, m_{\ell n}\}} \\
\hat{L}_\ell(n)_z \psi_{\{\rho_\ell, k_\ell, j_{\ell n}, m_{\ell n}\}} &= m_{\ell n} \psi_{\{\rho_\ell, k_\ell, j_{\ell n}, m_{\ell n}\}}
\end{align}
given explicitly by 
\begin{align}
\psi_{\{\rho_\ell, k_\ell, j_{\ell n}, m_{\ell n}\}}(\{h_\ell\}) = \prod_{\ell}
D^{(\rho_\ell, k_\ell)}_{j_{\ell\ell_+}\!m_{\ell\ell_+}
\,j_{\ell\ell_-}\!m_{\ell\ell_-}}(h_\ell)
\end{align}
where  $\ell_-$ and $\ell_+$ denote the source and target of $\ell$. These states are referred to as \textit{projected spin networks} 
\cite{EteraProj}
and form an orthonormal basis of the boundary Hilbert space, as follows from a version of the Peter-Weyl theorem as applied to $\SL^L$.

The transition amplitude for such a boundary state is then given by
\begin{align}
\label{projsn_amp_bf}
W_\mathcal{K}(\{\rho_\ell, k_\ell, j_{\ell n}, m_{\ell n}\} ) 
= \int \left(\prod_{\ell \in \gamma} dh_\ell \right) \psi_{\{\rho_\ell, k_\ell, j_{\ell n}, m_{\ell n}\}}(\{h_\ell\})W(\{h_\ell\}) .
\end{align}
By substituting in \eqref{discpi}, using the Plancharel theorem, inserting resolutions of the identity on each irrep $(\rho,k)$ of $\SL$ in terms of the states $|\rho, k; j, m\rangle$, and dropping one overall factor per vertex equalling the (divergent) volume of $\SL$,  one can show \cite{BaezIntro}.
\begin{align}
W(\{\rho_\ell, k_\ell, j_{\ell n}, m_{\ell n}\})
= \sum_{\{\rho_f, k_f, j_{fe}, m_{fe}\}}  
\left(\prod_{f \in \mathcal{K}} A_f\right) \left(\prod_{v \in \mathcal{K}}A_v\right)
\end{align}
where $A_f$ is a function only of $\{\rho_f, k_f, j_{fe}, m_{fe}\}_{e\in f}$,  and $A_v$ is a function only of $\{\rho_f, k_f, j_{f e}, m_{f e}\}_{e, f \ni v}$,  given explicitly by
\begin{align}
\label{BFvertex}
\begin{split}
A_v &= W_v\left(\{\rho_f, k_f, j_{f e}, m_{f e}\}_{e, f \ni v}\right) \\
& = \int \left(\prod_{e \ni v} dh_{e}\right)
\prod_{f \ni v}  D^{(\rho_f,k_f)}_{j_{fe}m_{fe}
\,j_{fe'}m_{fe'}}(h_{e}^{-1}h_{e'}) 
\end{split}
\end{align}
%
%
where $e,e'$ are the two edges sharing the same face $f$ at $v$, and where $\tilde{e}$ is an arbitrary choice of edge at $v$. 
Note that $A_v$ is just the BF transition amplitude \eqref{projsn_amp_bf}, for the projected spin network associated to the labels $\{\rho_f, k_f, j_{f e}, m_{f e}\}_{v\in e \in f}$, in the boundary Hilbert space of the single vertex $v$. The quantum numbers $\{\rho_f, k_f, j_{f e}, m_{fe}\}$ thus again have the interpretation of being quantum numbers of flux operators $\hat{J}_f(e)$, this time associated to each face $f$ in $\mathcal{K}$ and edge $e$ therein. While the flux operators on the boundary are parallel transported along each link, these flux operators in the bulk are parallel transported around each face, so that the quantum numbers $\rho_f, k_f$ depend only on $f$ and not $e$ \cite{Engle:2007qf}.  Eq. \eqref{BFvertex} is the vertex amplitude of the $\SL$ BF theory. By imposing simplicity on the boundary Hilbert space of each vertex, one obtains a proposal for the vertex amplitude of loop quantum gravity. 

\subsection{Quantum Simplicity}

Consider now a discretization of the linear primary simplicity constraints \eqref{KgL}, where each generator is associated with a pair node-link of each vertex graph:
\begin{align}
\label{discsimp}
\vec{S} := \vec{K}-\gamma \vec{L} \approx 0 .
\end{align}
We assume a simplicial decomposition with all tetrahedra space-like.
Due to the non-commutativity of the fluxes \eqref{Jpb}, part of these simplicity constraints become second class even without taking the secondary ones into account.
Dirac's original prescription for quantizing systems with constraints allows for only first class constraints to be implemented as operator equations in the quantum theory, as implementing second class constraints in this way would kill too many degrees of freedom and impose unphysical constraints. However, a unified method for imposing both first and second class constraints in quantum theory was later introduced by Thiemann --- the \textit{master constraint} method \cite{Dittrich:2004bp,Thiemann:2003zv} --- in which one combines the constraints into a choice of positive definite quadratic form, yielding a single constraint that vanishes if and only if all the original constraints vanish. In our case, $\SU(2)$-invariance restricts the quadratic form to be simply the square of $\vec{S}$,
\begin{align}
\label{clM}
M:= \vec{S}^2 = (1+\gamma^2)\vec{L}^2 - C_1 - 2\gamma C_2,
\end{align}
so that $M$ is a linear combination of the two $\SL$ Casimirs and the one Casimir $\vec{L}^2$ for the subgroup, $\SU(2)$. In quantizing the master constraint, one must allow for $\hslash$-scale modifications to ensure a sufficiently large space of solutions. This leads \eqref{clM} to be quantized as
\begin{align}
\nonumber
\hat{M}\ket{\rho,k;j,m} &:=
\left((1+\gamma^2)k^2 - (k^2-\rho^2) -2\rho k\right)\ket{\rho,k;j,m} \\
&=
\left((1+\g^2)(j^2-k^2) +(\rho-\g k)^2\right)\ket{\rho,k;j,m} .
\end{align}
Since both terms are non-negative definite,
setting this equal to zero implies both 
\be\label{gammasimple}
 k=j \qquad \text{and} \qquad \r=\g k.
\ee
We refer to these representations as $\g$-simple. A different convention with $-\g$ also appears in the literature, the relation being simply complex conjugation of all representations.
The map from the spin $j$ representation of $\SU(2)$ to the corresponding $\g$-simple representation 
$(\rho,k)=(\g j,j)$ of $\SL$ is called the $Y$-map in the early 
EPRL literature.\footnote{The difference with the BC model is that the latter imposes simplicity of $J$, as opposed to $\g$-simplicity. As a result, the only allowed irreps are either $\r=0$ or $k=0$, and are interpreted as describing a model with space-like or time-like faces respectively.}

The resulting matrix elements satisfy a sort of analytic property: they are fully determined by their restriction to $\SU(2)$, via an integral kernel defined in terms of $\SL$ and $\SU(2)$ characters \cite{Dupuis:2010jn,Rovelli:2010ed},
\be\nn
D^{(\r,j)}_{jmjn}(h) = \int_{\SU(2)} dg K(g,h) D^{(j)}_{mn}(g), \qquad K=\sum_j d_j^2\int_{\SU(2)}dk\chi^{(\r,j)}(gk)\chi^{(j)}(kh).
\ee
This property can also be understood in classical terms, recalling that these representations are a quantization of the canonical manifold $T^*\SL$. First, one can split the three equations \eqref{discsimp} into a first-class Lorentz-invariant constraint, given by
$\g(\vec L^2-\vec K^2)=(1-\g^2) \vec K\cdot \vec L,$
and two second class constraints (a nice explicit decomposition can be obtained working with spinors \cite{Speziale:2012nu}). The first involves the Casimirs and it is thus the same among left and right-invariant generators, whereas the second has to be implemented independently on both. This gives a total of 1 first class constraint and 4 second class constraints. It can be shown that symplectic reduction gives a 6d canonical submanifold isomorphic to $T^*\SU(2)$ \cite{Speziale:2012nu}, with holonomy $g$. 
The analytic property mentioned above is the quantum manifestation of this classical reduction.

There is another useful aspect that can be learned from this reduction.
Imposing the constraints but not dividing by the orbits, one has a 7d space whose extra dimension is spanned by a real variable $\Xi$ that spans each orbit of the  first class constraint. One can then construct a general $\SL$ holonomy $h(g,\Xi)$ in this space. In the absence of secondary constraints, we can treat $\Xi$ as gauge and remove it, for instance gauge-fixing it to 0 so to reduce trivially $h$ to $g$. If secondary constraints are present, their solution can always be seen as providing a specific gauge-fixing in terms of the reduced data --- namely, a non-trivial section $\Xi(g,X)$, the discrete analogue of the mapping $K=K(A,E)$ they provide in the continuum.
One can then argue that $g$ provides the correct discrete analogue of the Ashtekar-Barbero (AB) connection \cite{Speziale:2012nu}, whence the $\g$-simple matrices provide a quantum version of the map between $\SL$ holonomies and AB holonomies, provided the correct non-
trivial section is used.
These considerations can be useful to distinguish the role of the two
different connections present in the model, the spin connection and
the AB connection.

%

\subsection{Definition of the 4-simplex EPRL amplitude}

At this point we return to the quantum BF theory introduced in section \ref{sect:bf}, but now require that the cell complex $\mathcal{K}$ be dual to a simplicial complex. In particular, each vertex is now dual to a 4-simplex, and one can consider the BF amplitude \eqref{BFvertex} for a given set of data on the boundary of such a 4-simplex. By restricting this boundary data to satisfy the quantum simplicity constraints \eqref{gammasimple}, the boundary data reduce to those for a single vertex in loop quantum gravity \eqref{sfansatz}, and yield the \textit{EPRL vertex amplitude}:
\be
\label{eprlvert}
A_v(j_{f},m_{fe})= 
\int \left(\prod_{e} dh_{e}\right) \delta(h_{\tilde{e}})\prod_{f} D^{(\g j_f,j_f)}_{j_f m_{fe}\,j_{f}m_{e'f}}(h_{e}^{-1}h_{e'}),
\ee
where $\tilde{e}$ is an arbitrary choice of edge at $v$. 
One of the five group integrals over the non-compact manifold $\SL$ is redundant and has to be dropped to ensure finiteness of the amplitude \cite{Engle:2008ev}. This is achieved by the Dirac delta, which is with respect to the Haar measure $dh_{\tilde{e}}$ and peaked at the identity. The definition is valid only for $j_{f}\neq 0$. If a face has zero spin, the matrix should be replaced by 1.\footnote{This definition is necessary because $D^{(0,0)}(h)\neq 1$.}
The model is then completed taking trivial edge amplitudes, and $2j_f+1$ for the face amplitudes in \eqref{sfansatz}.
We first review results concerning a single vertex, and later on comment on multi-vertex amplitudes.  Multi-vertex analysis is crucial for understanding curvature, dynamics, and questions related to infrared divergences and causality; however most results of this analysis are not yet fully settled, so that we will mostly limit ourselves to point to the relevant research directions and literature.

When working with a single 4-simplex, it is often convenient to simplify the notation, and use $a=1,...5$ to label the edges, namely the nodes of the vertex graph $\G$, and $(ab)$ for the faces, namely the $10$ links between the nodes (we will no longer need these latin letters for hypersurface tensors). $\tilde{e}$ will then be chosen to correspond to $a=1$, so that the integrals \eqref{eprlvert} are only over $h_a$ with $a=2,...5$. In the following we will alternate between the two notations, with the dictionary just spelled out in mind.

Notice that the $Y$-map is imposed only at the tetrahedra: If we split the product $h_a^{-1}h_b$ using the group product, we obtain arbitrary weights:
\be
D^{(\g j,j)}_{jmjn}(h_{a}^{-1}h_{b}) = \sum_{l=j}^\infty\sum_{p=-l}^l D^{(\g j,j)}_{jmlp}(h_{a}^{-1})D^{(\g j,j)}_{lpjn}(h_{b}) .
\ee
If one imposes additional $Y$ maps on the internal weights, 
namely truncating the above sum over $l$ to its minimal value, 
one obtains a toy model, closely related to $\SU(2)$ BF theory and dubbed simplified EPRL model in \cite{Speziale:2016axj}, which is useful for numerical investigations with Euclidean boundary data.

\subsection{Properties}

{\bf Covariance.} The integral \eqref{eprlvert} defines (the lowest weights of) an $\SL$ invariant tensor, and can be expressed in terms of the corresponding Clebsch-Gordan coefficients.
Thanks to the $Y$-map, the quantum labels are in one-to-one correspondence with $\SU(2)$ weights, hence it can be used to provide transition amplitudes for LQG with 4-valent spin networks. Bulk Lorentz invariance and boundary covariance was proven in \cite{Rovelli:2010ed}.

Furthermore, the vertex amplitude is manifestly invariant under global $\SU(2)$ transformations at each edge. Therefore it can be contracted with invariant intertwiner tensors without any loss of information, giving the spin-intertwiner version
\be\label{A1int}
A_v(j_f,i_e) = \sum_{m_{fe}} 
{I}_{j_fm_{fe}}^{(i_e)}
A_v(j_f,m_{fe}).
\ee
The intertwiner label $i_e$ depends on a choice of recoupling basis, and $I$ is an orthonormal basis for intertwining tensors such as Wigner's 3jm symbols.

\smallskip\noindent {\bf Coherent states.}
The boundary magnetic indices can be superimposed to obtain $\SU(2)$ coherent states, as suggested initially in \cite{Livine:2007vk}. This allows a simple geometric interpretation of the amplitudes, and gives a useful handle to study the dynamics.
To define the coherent amplitude, we take lowest weight coherent states, 
labeled by $\vec n_{ab}$ at the source and $-\vec n_{ba}$
at the target of each link. 
 A standard calculation exploiting the factorization property of the coherent states leads to \cite{Livine:2007vk,BarrettLorAsymp,Dona:2019dkf}
\be\label{Ac}
A_v(j_{ab},\vec n_{ab}, -\vec n_{ba}) = e^{i \sum_{(ab)}j_{ab}\psi_{ab}}
\int \prod_{a=2}^N dh_a  \int \prod_{(ab)}  \f{ d\m(z_{ab})  }{ |\!|h_a^{\dagger}z_{ab}|\!|^2 |\!|h_b^{\dagger}z_{ab}|\!|^2} \exp S,
\ee
where the action is 
\be\label{action}
S(h,z,\z):=\sum_{(ab)} j_{ab} \ln \frac{\bra{\z_{ab}} h_{a}^{\dagger}z_{ab}\ra^{2} \la{h_{b}^{\dagger}z_{ab}}| {\z}_{ba}]^2}{|\!| h_{a}^{\dagger}z_{ab}|\!|^2\, |\!| h_{b}^{\dagger}z_{ab}|\!|^2 } +i\g j_{ab}\ln\frac{|\!|h_{b}^{\dagger}z_{ab}|\!|^2 }{|\!|h_{a}^{\dagger}z_{ab}|\!|^2 },
\ee
with $h_1=\Id$.
In this expression there are two types of spinors, latin and greek. 
The latin spinors $\ket{z_{ab}}$ provide the homogeneous realization of the infinite dimensional $\SL$ irreps. The integrand is invariant under complex rescalings $\ket{z_{ab}}\mapsto\l_{ab}\ket{z_{ab}}$, and the spinorial integration is defined over $\C P^1 \cong S^2$. 
See \cite{Ruhl,BarrettLorAsymp} for more details.
The Greek spinors are determined by the boundary normals via the map $(\vec n_{ab},-\vec n_{ba})\mapsto ( \ket{\z_{ab}},|\z_{ba}])$.
This map is not unique, and its choice determines the overall phase $\psi$. See the Appendix of \cite{Dona:2019dkf} for details.

An important property of the coherent amplitude is that its norm is invariant under rotations of the boundary data at each node. In other words, its norm depends only on rotational-invariant quantities such as the angles between the normals, whereas its phase depends also on the orientation of the normals.

\smallskip\noindent {\bf Representation in twistor space.}
The amplitude can be derived as an integral in twistor space \cite{Wieland:2011ru,Speziale:2012nu}. 
This is obtained through the parametrization of each copy of $T^*\SL$ in the phase space \eqref{cotbundle}
in terms of a pair of twistors with matching $\SL$ Casimirs, and it is useful to provide both a bridge to twistor theory, and a complementary derivation of the vertex amplitude from a discrete BF action. 
The twistorial description offered an independent insight into the flatness 
issue
\cite{Wieland:2013cr}, mathematical tools to describe time-like \cite{Rennert:2016rfp} and null \cite{Speziale:2013ifa} hypersurfaces in spinfoams, and potentially different implementations of the simplicity constraints along the lines suggested in \cite{Dupuis:2011fz,Dupuis:2011wy}. It also spurred a brief attempt at studying more the relation between the tools used in LQG and twistors \cite{Dunajski:2019yuy}.

\subsection{Large spin asymptotics}
\label{sect:asym}

The action \eqref{action} has two useful properties: its real part has upper bound 0, and it is linear in the spins. The latter means that the integral can be approximated using saddle point techniques in the limit of large spins. We describe only the case in which all spins are homogeneously large. Inhomogeneous limits have been studied for $\SU(2)$ amplitudes, see e.g. \cite{Khersonskii:1988krb,Bonzom:2009zd}.
The result of this analysis shows that the vertex amplitude decays exponentially, unless the boundary data satisfy some special configurations, and then the decay is power-law. The power-law decay requires boundary data such that the action admits a dominant saddle point, namely a point such that the gradient of the action with respect to all integration variables vanishes, and such that the real value of the action at the saddle point takes its maximal value. 
To understand the meaning of the special configurations, it is useful to interpret the boundary data in terms of discrete geometries. This can be done as follows. On each oriented link, we interpret the triple $(j_{ab}, \vec n_{ab}, \vec n_{ba})$ as the area and the normal directions of the triangle dual to the link in the frame of the tetrahedron dual to the source and target of the link. These variables parametrize the subset of the holonomy-flux phase space $T^*\SU(2)$ with vanishing twist angle $\xi_{ab}$ \cite{twigeo}.\footnote{The reason one considers boundary data with vanishing twist angles is that the amplitudes are coherent in the directions, but sharp in the areas. One can consider superpositions of vertex amplitudes also in the spins, like it is done in the propagator calculations \cite{Bianchi:2006uf}. These will be labelled also by the twist angles, and the boundary data then parametrize the complete holonomies and fluxes.}
A special feature of a flat 4-simplex is that the Jacobian between edge lengths and areas is generically invertible. Therefore, one can use the spins alone to determine a unique Euclidean 4-simplex (the spins satisfy the Euclidean triangle inequalities), a unique Lorentzian one (the spins satisfy the Lorentizian triangle inequalities), or no 4-simplex at all (no triangle inequalities are fully satisfied). Therefore whether the power-law behaviour exists depends entirely on whether the unit vectors are suitably adapted to the spins. To that end, we list some special subsets of the boundary data  which are relevant to the saddle point analysis.
\begin{enumerate}
\item[$(i)$] {\bf Twisted geometries.}\footnote{Or \emph{closed} twisted geometries, if the name twisted geometries is used for the generic parametrization of $T^*\SU(2)$ prior to imposing the closure conditions.} These are the data satisfying the closure conditions
\begin{equation}\label{Clos}
\sum_{b\neq a}j_{ab}\vec{n}_{ab}=0, \qquad \forall a.
\end{equation} 
Notice here the advantage of choosing the minus sign for the normals at the targets in \eqref{Ac}. This allows us to write the closure constraints above without further need of specifying orientations. These data describe a tetrahedron with areas given by $j_{ab}$ and 3d dihedral angles 
\be\label{defphi}
\vec n_{ab}\cdot\vec n_{ac} = \cos\phi^a_{bc}.
\ee
Looking at a single node, this information allows one to reconstruct the entire geometry of a flat tetrahedron: 4 areas and 2 independent dihedral angles among the 6 possible ones. This gauge-invariant information defines the shape of the tetrahedron. The collection of such tetrahedra defines a twisted geometry. The geometry is twisted because the shapes of the triangles don't necessarily match when computed in the source or target frames. It has on the other hand a well-defined notion of signature, depending on whether the (edge-dependent) 4d dihedral angles $\th^a_{bc}$ that can be constructed from the $\phi$'s using the spherical cosine laws satisfy Euclidean or Lorentzian inequalities (see e.g. \cite{Dona:2020yao}).

\smallskip

For these data there still is no dominant critical point, hence the amplitude decays exponentially.

\smallskip

\item[$(ii)$] {\bf Vector geometries.} 

These were introduced in \cite{Barrett:2002ur} as the set satisfying closure and additionally the orientation conditions
\be\label{orientations}
R_a \vec n_{ab} = - R_b \vec n_{ba}, \qquad R_a \in \SO(3).
\ee
A gauge-invariant description of vector geometries is not known. A partial answer in terms of four shape parameters and one non-gauge-invariant angle is given in \cite{Dona:2017dvf}. 
These conditions can be satisfied only for twisted geometries that have a Euclidean signature everywhere

\smallskip

If (and only if) the spins are Euclidean, these data admit a (unique) dominant saddle point. The amplitude then decays with the power law $j^{-12}$, and a phase that depends on the orientation of the data. 

\smallskip

\item[$(iii)$] {\bf Regge geometries.} 

These are data such that the shapes of the triangles match. In this case there is a well-defined notion of edge lengths of the triangulation, and these define a 3d Regge geometry. 
Furthermore, since every triangulation of $S^3$ by 5 tetrahedra is flat-embeddable, the same data also define a unique 4-simplex, which can be either Euclidean or Lorentzian. 
The shape-matching also guarantees that both the 3d dihedral angles and the 4d ones defined from the spherical cosine laws are edge-independent, hence we can write the relation between them as $\th_{ab}.(\phi)$.

\smallskip

For these data there are two distinct dominant saddle points. The amplitude then decays with the power law $j^{-12}$, a global phase that depends on the orientation of the data, and a relative phase which is proportional to the Regge action
\be
\label{SR}
S(j_{ab},\phi^a_{bc})=\g \sum_{ab} j_{ab}\th_{ab}(\phi).
\ee
Numerical investigations shows that the Hessians at the critical points are related by complex conjugation \cite{Dona:2019dkf}, therefore the relative phase can be written as a cosine.

\end{enumerate}
From the interpretation of the critical points one can also conveniently group these types of data into `degenerate'(vector and Euclidean Regge) and `non-degenerate' (Lorentzian Regge) (see section \ref{sectprop}).

The explicit appearance of the Regge action in the large spin limit is particularly interesting. It is in fact known that the Regge action has a well-defined continuum limit given by the Einstein-Hilbert action, so that this result might be understood as a first step in proving   that the model reproduces general relativity. However, the situation turns out to be more subtle.

One can extend the saddle point approximation to include the sums over the spins associated with an internal face, and this has been shown to imply that, in the limit of large boundary spins, the theory on a fixed cell-complex is dominated by flat solutions only 
\cite{Bonzom:2009hw,Hellmann:2012kz,Perini:2012nd,Han:2013hna,Hellmann:2013gva,Engle:2020ffj,Gozzini:2021kbt}.
This shows that general relativity cannot be recovered from the theory by first taking the large spin limit and then taking the continuum limit. Indeed, from simple familiar examples one might have expected this \cite{Engle:2021xfs}. Rather, one must take both limits at the same time, while ensuring a certain model-dependent inequality involving the curvature, the spin, and the Immirzi parameter holds in the process \cite{Han:2017xwo,Asante:2020qpa,Asante:2020iwm,Han:2021kll}. 
While this clarifies the way in which the classical limit must be taken, whether the correct classical limit can be obtained in this way with sufficient generality remains an open question.

\subsection{Booster decomposition and numerical evaluations}

Numerical evaluations of \eqref{eprlvert} are difficult. It is a multidimensional unbounded integral over ratios of sums of hypergeometric functions which not only require high numerical accuracy but are furthermore rapidly oscillating. Furthermore, the results need to be tabulated for a large number of magnetic number combinations. It is significantly harder than numerics for the Barrett-Crane or $\SU(2)$-based models.

One method to tackle this challenge is to exploit the decomposition of $\SL$ Clebsch-Gordan (CG) coefficients into $\SU(2)$ ones \cite{Speziale:2016axj}.
This allows one to rewrite the vertex amplitude \eqref{A1int} as a sum over $\SU(2)$ 15j symbols labelled by new virtual spins and interwiners to be summed over:
\be
A_v(j_f,i_e)= \sum_{l_f=j_f}^\infty\sum_{k_e} \{15j\}_{l_f,k_e} \prod_{e=2}^5 B_4^{\g}(l_f,k_e;j_f,i_e).
\ee
The sums over the intertwiner labels $k_e$ range over the usual CG inequalities, and 
\be
B:=\sum_{m_f=-j_f}^{j_f} \vet{l_f}{m_f}^{(k_e)} \vet{j_f}{m_f}^{(i_e)}  \int_o^\infty dr \f{\sinh^2r}{4\pi}\, \prod_{f\in e} d^{(\g j_f,j_f)}_{l_fj_fm_f}(r)
\ee
are called `booster' functions. The integrands  $d^{(\r,k)}$ are the reduced boost matrix elements \cite{Ruhl}.

The advantage of this method is that the unbounded integrals have been reduced to a single one per edge. Furthermore, the vertex amplitude is the much simpler 
$\SU(2)$ one, very fast to evaluate numerically. The price to pay is the introduction of new internal sums, which are unbounded, and have to be truncated by hand in any numerical evaluation. The sum converges, but the speed of convergence depends on the configuration considered, and it is typically slower for Lorentzian boundary data, as opposed to Euclidean ones. Writing and optimizing numerical codes performing this numerical evaluation is a current focus of research
\cite{Dona:2018nev,Gozzini:2021kbt,Dona:2022dxs,Dona:2022yyn}.

The booster decomposition is not the only approach to numerical evaluations: there exist by now a rich set of ideas and methods that have been developed to compute different aspects of the EPRL amplitude, and more in general of spinfoam models, see e.g. 
\cite{Bahr:2015gxa,Bahr:2017klw,Dona:2017dvf,Han:2020npv,Han:2021kll,Asante:2020qpa,Allen:2022unb}.
The Chapter ``Spinfoams and high performance computing'' in this Handbook \cite{Dona:2022yyn} is dedicated to some of these methods.

\subsection{Extension to non-simplicial spinfoams: the KKL model}

If one wants to give amplitudes to all possible spin network states, the first step is to define the model beyond 4-simplices. A nice way to systematically construct non-simplicial vertex amplitudes was studied in \cite{Kaminski:2009fm}. It was there applied to a straightforward extension of the EPRL model, based on applying the same set of constraints \eqref{gammasimple} to every link in the generalized vertex graph. The resulting amplitude which we refer to as KKL model, is the immediate generalization of \eqref{eprlvert} with the connectivity in the arguments of the representation matrices induced by the vertex graph.

This generalization should, however, be taken with care, because it may be ill-defined depending on the graph. The convergence of the non-compact group integrals is not guaranteed by simply removing one redundant integration 
as in \eqref{eprlvert}, and depends on the connectivity of the graph. A sufficient condition is 3-link-connectivity, namely any bi-partition of the nodes cannot be disjointed by cutting only two links, but many non-3-link-connected graphs are well defined, see \cite{Baez:2001fh,Kaminski:2010qb,Sarno:2018ses}. This condition is satisfied by the 4-simplex graph.

The KKL model is appealing for its simplicity. Recall, however, that beyond the 4-simplex, the linear simplicity constraints are no longer sufficient to recover a discrete flat geometry around one vertex, already at the classical level. One may then expect the model to be dominated by a more general class of geometries than the Regge geometries appearing in the 4-simplex case. This is indeed what happens, and interestingly, the generalized geometries that appear have both a simple geometric description, and admit a simple generalization of the Regge action \cite{Bahr:2015gxa,Bahr:2017klw,Dona:2017dvf,Dona:2020yao}.

This is explained observing that the conditions required to obtain a second distinct saddle point are angle-matchings: they are equivalent to shape-matchings because once the area of two triangles match, matching the angles identifies completely the triangles up to rotations. But matching the area and the angles of two $n$-sided polygons still leaves $n-3$ unmatched variables, corresponding to conformal transformations of the polygons. These \emph{conformal twisted geometries} describe the critical behaviour of the KKL model with distinct critical points. And the reason why the Regge action \eqref{SR} is still well-defined and indeed arises at the saddle points, is that angle-matching is enough to have edge-independence of the 4d dihedral angles defined from the spherical cosine laws \cite{Dona:2020yao}. 

Applications of the KKL model include \cite{Bianchi:2010zs,Vidotto:2011qa,Sarno:2018ses}.
An attempt to modify the KKL model to include more constraints appeared in \cite{Assanioussi:2020fml}.

\subsection{Proper vertex}
\label{sectprop}

Consider a single 4-simplex $v^*$, with coherent boundary data $j_{ab}=j_{ba}, \vec{n}_{ab}$ for $a,b\in\{1,\dots, 5\}$. The expression for the vertex amplitude consists in an integral over the five $\SL$ group elements $\{h_a\}$. Introduce an arbitrary flat connection $\partial$ inside the 4-simplex, such that the 4-simplex is the convex hull of its 0-simplices.
When the critical point equations of the group integrals are satisfied, there exists a unique lie algebra valued 2-form $B^{IJ}$, constant with respect to $\partial$, such that 
\begin{align}
\int_{\Delta_{ab}} B^{IJ} =  h_a  \triangleright \left( j_{ab} (1,0,0,0) \wedge (0,\vec{n}_{ab})\right)^{IJ}
\end{align}
%
%
for all $a,b$, where $\Delta_{ab}$ denotes the triangle in $v^*$ between tetrahedra $a$ and $b$, oriented as part of the boundary of $a$, and $\triangleright$ denotes the adjoint action \cite{Engle:2015mra}. 
The resulting two form $B$ is either degenerate ($\tr (B \wedge \star B) = 0$) or is of the form $(I\pm)$ in \eqref{sectors} for some tetrad $e$ in $v^*$, constant with respect to $\partial$ \cite{Engle:2015mra}. This is the precise sense in which the linear simplicity constraint, as imposed in the EPRL model, eliminates the sectors $(II\pm)$ in \eqref{sectors}. 
For boundary data corresponding to a vector geometry or Euclidean Regge geometry,
the critical point(s) selected corresponds to a two form $B$ that is degenerate, whereas for data corresponding to a Lorentzian Regge geometry, the two critical points selected, corresponding to the two sectors $(I\pm)$, giving rise precisely to the two exponentials in the large spin asymptotics equaling the cosine of the Regge action
\cite{BarrettLorAsymp}. 
The existence of these multiple sectors are potentially problematic for a number of reasons:
\begin{enumerate}
\item The degenerate sector, which classically must be excluded to recover general relativity, is present and not suppressed.
\item Even if the degenerate sector were eliminated, the existence of two different non-degenerate sectors is potentially problematic, because there are no critical point equations forcing the sector to be the same across different 4-simplices. As a consequence, the actions appearing in the large spin limit of the amplitude, for sufficiently large triangulations, include sign flips as one goes from 4-simplex to 4-simplex, which changes the corresponding equations of motion so that they no longer approximate those of general relativity. This contrasts with the classical theory, where elimination of the degenerate sector is enough to ensure that the solution is either entirely in the sector $(I+)$ or entirely in the sector $(I-)$.\footnote{That being said, there are choices of triangulation and boundary data that force the sector to be uniform, relaxing this issue in such cases \cite{Han:2021kll}.}
\item Results from the analysis of 3d gravity \cite{Christodoulou:2012af} suggest that such non-gravitational equations of motion, by allowing for unsuppressed ``spikes'', are furthermore responsible for bubble divergences in spinfoam models with zero cosmological constant, such as EPRL. Such divergences have so far impeded calculations involving sufficiently refined triangulations.
\end{enumerate}
In addition, it has been suggested that these multiple sectors are an obstruction to an exact match with the canonical theory 
\cite{Thiemann:2013lka}, and that restriction to a single sector may be analogous to the positive frequency condition in loop quantum cosmology necessary to extract correct physics \cite{Ashtekar:2010ve}. A similar restriction has also been advocated by Oriti \cite{Oriti:2004mu} as a way of implementing causality in the sense introduced by Teitelboim \cite{Teitelboim:1981ua}.

For these reasons, it is interesting to see if it is possible to remove the degenerate and $(II-)$ sectors from the vertex amplitude, so that only the $(II+)$ sector remains. This is the idea behind the ``proper vertex'', in which a classical inequality is first derived selecting the $(II+)$ sector, which then is quantized as a projector and inserted into the EPRL vertex amplitude
\cite{Engle:2012yg, Engle:2015mra}. Results so far for the proper vertex reproduce the successful tests of the EPRL vertex amplitude \cite{ChaharsoughShirazi:2015uuu, Vilensky:2016tnw}, as well as providing indications that some of the above issues might be resolved \cite{Engle:2015zqa, Engle:2011un}. One drawback of the proper vertex is that it is less simple than the ERPL vertex, suggesting that it may be worthwhile to seek a simpler alternative that achieves the same goals.

\subsection{Extension to non-space-like building blocks}
\label{othersig}

Within the framework of Regge calculus, it is possible to approximate arbitrarily well a pseudo-Riemannian geometry using space-like tetrahedra, and this is the reason why most models are built this way. However, it is also true that allowing 4-simplices with non-space-like faces and tetrahedra allows a richer sampling of the continuum spacetime, as well as more symmetric choices of shapes for simple geometries. With such motivations in mind, it is possible to consider a version of the  BC and EPRL models with not only space-like but also time-like faces, see e.g. \cite{Barrett:1999qw,Conrady:2010kc,Kaminski:2017eew,Han:2021bln}. A particularly interesting result in this generalization is that a 4-simplex amplitude with mixed faces does not admit vector geometries as critical points, hence its semiclassical limit is dominated by Regge geometries alone \cite{Liu:2018gfc}. Using null-faces on the other hand remains more elusive \cite{Neiman:2012fu,Speziale:2013ifa,Speziale:2014uua,Jercher:2022mky}.

\subsection{Inclusion of cosmological constant}

The basic idea of the EPRL model, and other spinfoam models before it, is to start from $\SL$ quantum BF theory on a simplicial complex, and then impose the simplicity constraints at the tetrahedra in order to obtain gravity, imitating Plebanski at the quantum level. Besides boundary labels being restricted by simplicity, the vertex amplitude of EPRL is simply the vertex amplitude of BF theory, and the flatness equation of motion of BF theory \eqref{bfflat} is the reason why each 4-simplex is flat in the large spin limit. The most recent and well-developed extension of EPRL to include a cosmological constant 
\cite{Haggard:2014xoa, haggard2016four, haggard2016encoding, Han:2021tzw} builds on this basic approach, except that, instead of starting from BF theory, it starts from a modification of BF theory, which one may call ``Holst-$\Lambda$BF'', that includes the cosmological constant in a way that also involves the Immirzi parameter. Specifically the action is that in equation \eqref{S} with the Lagrange multiplier $\phi$ removed. 
When the form of $B$ in the $(I\pm)$ sectors of the solution to simplicity are substituted into this action, it reduces to the Holst action of gravity with Newton constant $\pm G$ and Immirzi parameter $\gamma$, with an extra term giving cosmological constant $\pm \Lambda$, where the signs depend on the sector. Variation of the above action with respect to $B$ yields the equation of motion 
\begin{align}
\label{LBFconstcurv}
F = \frac{\Lambda}{3} B
\end{align}
so that the solutions are no longer flat as in \eqref{bfflat}, but of ``constant curvature'' in the sense above. In sectors $(I\pm)$, \eqref{LBFconstcurv} imposes constant Ricci scalar curvature $\pm 4 \Lambda$.

The lack of flatness is also reflected in the continuum path integral. Since the action is no longer linear in $B$, but quadratic, one no longer obtains a Dirac delta function imposing flatness as in \eqref{eq:contpi}, but rather a Gaussian integral, leaving as integrand the exponential of an action with Lagrangian quadratic in $F$ equaling a total derivative, so that the action reduces to an action on the 3d boundary, specifically the Chern-Simons action.
The level of the Chern-Simons theory is complex, given by $\kappa = \frac{12\pi}{8\pi \Lambda \ell_P^2 }\left(\frac{1}{\gamma} + i\right)$, with $\ell_P^2:= G\hslash$ the Planck area.
Invariance of the above expression under large gauge transformations forces $\Re \kappa \in \mathbb{Z}$, which forces the cosmological horizon area, $12\pi/|\Lambda|$, to be an integer multiple of $8 \pi \gamma G \hslash$ --- that is, that it correspond to a loop quantum gravity area eigenvalue with integer spin $\Re \kappa$.

The Chern-Simons theory on the boundary of each 4-simplex is topological, and so yields a finite number of degrees of freedom at each vertex. By excluding the links of the boundary graph from the theory, the resulting quantum degrees of freedom again match those of loop quantum gravity on the boundary graph, except with spins $j$ bounded above by $\Re \kappa/2 \in \mathbb{Z}/2$.



As one can see, this extension has many beautiful surprises, and leads to a completely new way to look at the theory. Moreover, the fact that it includes a cosmological constant is consistent with what we observe in our universe so far. It is interesting, and not necessarily problematic, 
that both signs of the cosmological constant are included in the theory. 

The fact that one obtains an upper bound on the spins is reminiscent of how quantum deformations of classical Lie groups can make the number of irreps finite. 
However, since the level $\kappa$ is complex, no relation of the above model to a quantum group has yet been established. Studies of 3d gravity have long suggested a relation between the inclusion of a cosmological constant and the use of quantum groups, starting from the Turaev-Viro model \cite{Turaev:1992hq}. See, for example, \cite{Perez:2010pm} and references therein. Attempts to use quantum groups to incorporate a cosmological constant into 4d spinfoam models also include \cite{Noui:2002ag,Han:2010pz,Fairbairn:2010cp}.


\section{Conclusions}
%
%

The spinfoam formalism provides a specific framework to define the path integral for quantum gravity so that its boundary states are the spin networks of the Hilbert space of loop quantum gravity. The goal of the program is to find a spinfoam amplitude that implements correctly the diffeomorphism constraints and reduces to general relativity in the appropriate limit. This would provide an ultraviolet complete quantum theory of gravity, whose geometrical and physical interpretation can be studied using results from loop quantum gravity on geometric operators.

Currently the most studied model is the EPRL model, of which we described here in some details the Lorentzian version.
This model has various promising properties: it provides transition amplitudes to all spin network states, it includes a non-trivial dependence on $\g$, it passes a test on 4-simplex correlations failed by the BC model, and has a clear relation to the Regge action in the large spin limit.
This last result has been often invoked as a first step towards a proof that the model reproduces general relativity, since it is known that the Regge action reduces to the Einstein-Hilbert action in the continuum limit. 
However, the situation is more complicated. As noted at the end of section \ref{sect:asym}, the large spin limit of the theory on a fixed cell complex is unrelated to general relativity, so that this limit must be combined appropriately with a refinement limit. 
Moreover, as discussed in sections \ref{sect:asym} and \ref{sectprop},
the Regge action appears in the large spin limit of the amplitude only for non-degenerate configurations, 
and there exist degenerate configurations whose amplitude has similar fall-off and so are not suppressed. Furthermore, even for non-degenerate configurations, the Regge action appears twice, once for each sign, independently for each 4-simplex, leading to actions for larger triangulations with alternating signs and no relation to gravity.
If these latter features turn out to indeed prevent the correct classical limit of the model, the proper vertex discussed in section \ref{sectprop} offers one possible modification to remove them. 
Beyond the key question of the semiclassical limit, numerous applications of the EPRL model have appeared in the literature, including the graviton propagator \cite{Bianchi:2011hp,Riello:2013bzw,Dona:2022vyh}, quantum tunnelling for the black-hole-to-white-hole transition \cite{Christodoulou:2016vny,Soltani:2021zmv}, and spinfoam cosmology \cite{Bianchi:2010zs,Ashtekar:2010ve,Henderson:2010qd,Livine:2011up,Vidotto:2011qa,Sarno:2018ses}.

In this chapter, we focused on the EPRL model, in 3+1 dimensions and the BF-plus-constraints approach, but in closing we would like to stress and convey that the spinfoam formalism provides a valuable framework for studying covariant transition amplitudes of quantum geometry more generally. The conceptual and technical framework we reviewed here can be applied and extended to models with different signature, different dimensions, or including matter sectors. It provides an arena for studying renormalization and continuum limit in a background-independent manner.

\subsection{Open questions and research directions}

Among the open question and research directions that are being investigated, we can mention the following ones.
\bit
\item 
In three dimensions, gravity is a topological field theory, and the only degrees of freedom appear on the boundary. The corresponding spinfoam construction gives the Ponzano-Regge and Turaev-Viro models, depending on the value of cosmological constant. The inclusion of boundary degrees of freedom in these models has been recently investigated, and led to a promising convergence of results with other approaches to quantum gravity \cite{Dittrich:2018xuk,Goeller:2019zpz}.
\item 
Explicit couplings to various types of matter have been written down \cite{Mikovic:2001xi,Speziale:2007mt, Fairbairn:2006dn,Bianchi:2010bn,Kisielowski:2018oiv}, but there is still work to do to understand the resulting matter dynamics, for instance whether one can see explicitly the regularization of UV divergences as has been argued for from the canonical perspective \cite{Thiemann:1996aw}. 
\item 
Spinfoam models are ultraviolet finite, but the EPRL model and its predecessors do have infrared divergences, in particular, so called `bubble divergences' that prevent calculation for sufficiently large triangulations.
From studies of 3d gravity, it has been suggested that such divergences are due to diffeomorphism invariance \cite{Freidel:2004vi}. However, on closer inspection, the configurations leading to the divergence are not diffeomorphism related, but correspond to ``spikes'' with non-zero deficit angles that are not isometric to any region in $\mathbb{R}^3$, so that they represent violations of the classical equations of motion, a violation which can be traced to the independent summing over orientations in each tetrahedron, the ``cosine problem''\cite{Christodoulou:2012af}. The proper vertex discussed above provides one possible solution to this.
When the cosmological constant is included, all sums are finite merely because the range of spins is finite, but it is not clear whether the underlying physical problem remains. 
\item 
The derivation of the spin-foam sum from the projector onto solutions of the Hamiltonian constraint (see section \ref{sect:sfansatz}) leads to a sum over all 2-complexes.
An open question is whether such a sum over 2-complexes is equivalent to considering a fixed 2-complex and then taking its refinement limit. Work is in progress to understand how the amplitudes scale when refining the cell complex \cite{Bahr:2017klw,Sarno:2018ses,Asante:2022dnj}.
\item 
As reviewed in section \ref{sect:sfansatz}, a motivation for the spinfoam framework comes from expanding the projector onto solutions of the Hamiltonian constraint operator. However, this derivation is incomplete, and the spinfoam ansatz is not exactly taken from it. As a consequence, the relation of the EPRL and other spinfoam models to canonical quantizations of the Hamiltonian constraint is an open issue. Work in this direction include, for example, \cite{Alesci:2013kpa,Thiemann:2013lka,Han:2020chr}.
\item The KKL extension on an arbitrary vertex lacks some of the classical simplicity constraints, and it is an open question how to add them, and whether they successfully reduce the dominant contributions from conformal twisted geometries to Regge geometries
\cite{Bahr:2015gxa,Bahr:2017klw,Dona:2017dvf,Dona:2020yao,Assanioussi:2020fml}.
\item Even on a single 4-simplex, the imposition of the primary simplicity constraints alone may fail to take into account the modification to the quantum measure that originates from the standard treatment of quantizing systems with second class constraints \cite{Alexandrov:2008da,Alexandrov:2010pg,Alexandrov:2012pj}.
\item Conversely, it has been argued in \cite{Asante:2020iwm} that including second class constraints in a quantum system with spacetime discreteness makes the emergence of semiclassical physics more subtle than simply taking the $\hslash\rightarrow 0$ limit. This observation has led to the construction of effective spin foam models based on the imposition of simplicity constraints in terms of discrete geometries, as opposed to a restriction of the quantum labels. 
The effective models offer a handle on the interplay between refining and the emergence of curvature, and also a useful link to area-Regge calculus with a new perspective on the recovery of general relativity \cite{Dittrich:2022yoo}.
\item The BF quantization used as starting point to impose the simplicity constraints depends only on data on a 2-complex. This has been called into question in \cite{Bonzom:2011br}, arguing that additional data on 3-cells is needed on manifolds of non-trivial topology.
\eit

\subsection{Take home message}

The foundations of the spinfoam formalism have multiple roots. From implementing the constraints via a path integral formulation, generalizing the mathematical framework of topological quantum field theories, and providing a sampling of quantum geometries, numerous ideas have converged towards an ansatz like \eqref{sfansatz}.
Explicit realizations of this idea have been constructed in the literature, and spurred many mathematical developments and applications. 
The current open questions and research directions offer a stimulating window on all facets of the quantum gravity problem, and this provides one of the key interests to keep working in the spinfoam approach.

\section*{Acknowledgements}

JE was supported in part by NSF Grant PHY-2110234.
SS was supported in part by Grant 62312 from the John
Templeton Foundation, as part of the
\href{https://www.templeton.org/grant/the-quantum-information-structure-of-spacetime-qiss-second-phase}{‘The Quantum Information Structure of Spacetime’ Project (QISS)}. 
The opinions expressed in this publication are those of the authors and do
not necessarily reflect the views of the John Templeton Foundation.

%

\providecommand{\href}[2]{#2}
\begingroup\raggedright
\renewcommand{\addcontentsline}[3]{}
\endgroup

\end{document}